\documentclass[11pt]{article}
\usepackage{graphicx} 
\usepackage[algo2e,algoruled,linesnumbered]{algorithm2e}
\usepackage{amsfonts}       
\usepackage{amsmath}
\usepackage{amssymb}
\usepackage{amsthm}
\usepackage{bbm}
\usepackage{bm}
\usepackage{color}
\usepackage{booktabs}       
\usepackage{caption}
\usepackage{graphicx}
\usepackage{hyperref}       
\usepackage[utf8]{inputenc} 
\usepackage{mathtools}
\usepackage{microtype}      
\usepackage{multirow}
\usepackage{newtxmath}      
\usepackage{nicefrac}       
\usepackage{thm-restate}
\usepackage{algorithm}
\usepackage{tikz}
\usepackage{url}            
\usepackage{subcaption}     
\usepackage{wrapfig}        
\usepackage{caption}
\usepackage{fullpage}
\usepackage[noend]{algpseudocode}
\newtheorem{theorem}{Theorem}

\newtheorem{lemma}[theorem]{Lemma}

\newtheorem{definition}{Definition}

\usepackage{todonotes}

\newcommand{\Alg}{\operatorname{Alg}}

\renewcommand{\paragraph}[1]{\vspace{0.2 cm} \noindent \textbf{#1} }

\newcommand{\bu}{\mathbf{u}}
\newcommand{\be}{\mathbf{e}}

\newcommand{\Vol}{\text{Vol}}
\newcommand{\CW}{{{Minimum Cutwidth}}\xspace}
\newcommand{\ST}{{{Minimum Storage}}\xspace}
\newcommand{\PW}{{{Minimum Pathwidth}}\xspace}
\newcommand{\LA}{{{Minimum Linear Arrangement}}\xspace}

\newcommand{\CIG}{{{Minimum Containing Interval Graph}}\xspace}
\newcommand{\STP}{{{Minimum Storage Time Product}}\xspace}

\newcommand{\RS}{{{Minimum Register Sufficiency}}\xspace}
\newcommand{\US}{{{Uniprocessor Scheduling}}\xspace}
\newcommand{\FAS}{{{Minimum Feedback Arcset}}\xspace}
\newcommand{\BW}{{{Minimum Bandwidth}}\xspace}
\newcommand{\PMM}{{{Peak Memory Minimization}}\xspace}
\newcommand{\CLA}{{{Minimum Cut Linear Arrangement}}\xspace}

\SetKwComment{Comment}{/* }{ */}

\newtheorem{question}{Open Question}

\title{On Approximating Cutwidth and Pathwidth}
\author{ Nikhil Bansal\thanks{\texttt{bansaln@umich.edu}. \small University of Michigan. Supported in part by  NWO VICI grant 639.023.812 and NSF award CCF-2327011.} \and
Dor Katzelnick\thanks{\texttt{dkatzelnick@cs.technion.ac.il}. \small Technion - Israel Institute of Technology.}  
\and Roy Schwartz \thanks{\texttt{schwartz@cs.technion.ac.il}. \small Technion - Israel Institute of Technology.
Research supported by European Union’s Horizon 2020 research and innovation program under grant agreement no. 852870-ERC-SUBMODULAR}
}
\begin{document}
\date{}
\maketitle

\begin{abstract}
We study graph ordering problems with a min-max objective.
A classical problem of this type is cutwidth, where given a graph we want to order its vertices such that the number of edges crossing any point is minimized. 
We give a $ \log^{1+o(1)}(n)$ 
approximation for the problem, substantially improving upon the previous poly-logarithmic guarantees based on the standard recursive balanced partitioning approach of Leighton and Rao (FOCS'88).
Our key idea is a new metric decomposition procedure that is suitable for handling min-max objectives, which could be of independent interest. We also use this to show other results, including an improved $  \log^{1+o(1)}(n)$ 
approximation for computing the pathwidth of a graph.

\end{abstract}
\allowdisplaybreaks

\section{Introduction}\label{sec:introduction}

We study graph layout problems where given a graph $G=(V,E)$ the goal is to find a linear ordering of the vertices $V$ 
that minimizes a certain objective.
Such  problems have been studied extensively in  algorithm design and graph theory and they arise naturally in diverse areas such as VLSI design, scheduling, graph minor theory, graph drawing and computational linguistics. 

These problems typically have either a {\em min-sum} or a {\em min-max} objective, where the goal is to minimize some average or some worst measure of the layout.
Some classical problems with {\em min-sum} objectives include, e.g., 
 \LA,  \FAS, \STP and \CIG
\cite{ LR88, ravi1991ordering, Hansen89,seymour1995packing,  even2000divide,rao2005new, feige2007improved, charikar2010ℓ}.

Some classical problems with {\em min-max} objectives include
\CW \cite{yannakakis1985polynomial,korach1993tree,LR88,AFK02,BFK12,WAT14} (a.k.a. \CLA), \PW \cite{Bod93,BGHK95,CDFKMS19,feige2005improved}, \RS \cite{klein1990approximation}, \BW 
\cite{BKRV00,FEIGE2000510,Gup01}, and  \ST \cite{kayaaslan2018scheduling,liu1987application} (a.k.a. \PMM).

\paragraph{Recursive Partitioning.}
In their seminal work, Leighton and Rao \cite{LR88} gave a generic approach 
for layout problems, based on recursive partitioning. The idea, dating back to \cite{BL84},
is to apply balanced cuts recursively until each piece becomes a single vertex. Using their $O(\log n)$ approximation for balanced cut, with another $O(\log n)$ factor loss due to recursion, 
\cite{LR88} and  \cite{Hansen89,ravi1991ordering} gave $O(\log^2 n)$ approximation for many of the problems  above (with the notable exception of \BW which
differs from other layout problems and requires other techniques).
We refer to the journal version \cite{LR99} for more applications of this idea.

For many of the problems mentioned above the bound improves to $O(\log^{3/2}{n})$ by directly
plugging the breakthrough $O(\log^{1/2}n)$ approximation for balanced cuts of Arora, Rao and Vazirani \cite{arora2009expander} based on SDP relaxations.

\paragraph{Improvements for Min-Sum Objectives.}
Following the work of \cite{LR88}, there has been impressive progress on problems with min-sum objectives based on several ingenious ideas.

Seymour \cite{seymour1995packing} gave a beautiful way to trade off the quality of a cut with its imbalance, and improve the approximation for \FAS to $O(\log n\log \log n)$.
This was extended further by Even et al., \cite{even2000divide}, using the idea of spreading metrics, to obtain $O(\log n \log \log n)$ approximation for all the min-sum problems above and more.
Rao and Richa \cite{rao2005new} gave another approach and improved the bounds for \LA and \CIG to $ O(\log{n})$.
Even more remarkably, Charikar et al., \cite{charikar2010ℓ}, and Feige and Lee \cite{feige2007improved} combined these ideas with the 
SDP based techniques of \cite{arora2009expander},  
to obtain $ O(\log^{1/2}{n}\log{\log{n}})$ approximations for various $\min$-sum problems mentioned above. 

Taking a slightly broader perspective, recursive partitioning and its variants have become basic tools in algorithm design and have also led to impressive progress in several related areas such as metric embeddings \cite{KLMN05, ALN07} and tree embeddings \cite{Bartal96, Bartal98,FRT04}, oblivious routing \cite{R08, R09}, expander decompositions\cite{GT14,KVV04,ST11} and distributed computing \cite{AP90}. As this literature is extensive, we do not discuss it at all here.

\paragraph{Min-Max Objectives.} Surprisingly however, there have been no improvements for layout problems with min-max objectives beyond recursive partitioning. 
Roughly, min-max objectives are harder to handle than corresponding min-sum objectives as they require controlling mutliple quantities at once. For this reason, it is unclear how to use here the various techniques developed for min-sum objectives. We discuss this issue further in Sections \ref{sec:overview}  and \ref{sec:discussion}.

In particular, despite much interest, the best known bounds for various central problems such as \CW and \PW 
are still $O(\log^{3/2} n)$ (and $O(\log^2 n)$ if we restrict to LP-based methods), based on the approach of Leighton and Rao \cite{LR88}.

\subsection{Our Results}\label{sec:results}

We give the first improved approximation bounds for \CW and \PW, beyond the recursive partitioning framework. 
These are the most well-studied and fundamental $\min$-$\max$ layout problems on undirected graphs. 
We now state these problems and our results.

\paragraph{Minimum Cutwidth.}
In the \CW problem, we are 
given an undirected graph $G$ on $n$ vertices, and the goal is to order its vertices to minimize the {maximum} number of edges crossing any point.\footnote{This problem is also referred to as Minimum Cut Linear Arrangement (but we prefer to use \CW to avoid confusion with the \LA problem).} 

Formally, given an ordering (permutation) $\pi: V \rightarrow [n]$, 
let $S^\pi_i\triangleq \{v:\pi(v)\leq i\}$ be the set of the first $i$ vertices in the ordering. 
Let us define $\mathrm{CW}_{\pi}(G)\triangleq \max _{i\in [n]}|\delta(S^{\pi}_i)|$,  where  $\delta(S^\pi_i)$ is the cut $ E(S^\pi_i,V\setminus S_i^\pi)$. Then, the cutwidth of $G$ is
\[
    \mathrm{CW}(G)\triangleq \min_\pi \mathrm{CW}_{\pi }(G)  , \label{def:CW}
\]
where the minimum is taken over all permutations $\pi$ of $V$. 

The min-sum version of cutwidth is the \LA (MLA) problem. Here,  
 we want to find an ordering $\pi$  that minimizes
$\sum_{i=1}^n |\delta(S^\pi_i)|$, or equivalently
the {total} edge length $\sum_{(u,v)\in E} |\pi(u)-\pi(v)|$.

We show the following result for \CW.
\begin{theorem}\label{thrm:CW}
There is an efficient $ O(\beta(n) \log{n}) $ approximation algorithm for \CW, where $\beta(n) =\exp(O(\sqrt{\log{\log{n}}})) = \log^{o(1)}(n)$.
\end{theorem}
As stated above, the best previous result was based on generic recursive partitioning (so $O(\log^2 n)$ approximation using LP-based techniques due to Leighton and Rao \cite{LR88}, and $O(\log^{3/2}n)$ approximation using the $O(\log^{1/2} n)$-approximation for balanced cut due to Arora et al. \cite{arora2009expander}).

Our result is LP-based, and it is well-known (based on standard integrality gaps on expanders), that $O(\log n)$ is the limit of these methods. 
Our main technical contribution is a new metric decomposition method that allows to simultaneously control multiple cuts (which is necessary for min-max objectives). We give an overview in Section \ref{sec:overview} and describe the details in Section \ref{sec:cutwidth}. 

Interestingly, our decomposition  method is also quite robust. In particular, it also works for $\min$-sum objectives (albeit with an extra loss of $\beta(n)$ compared to known approaches) and also extends to vertex separator variants. We describe these results next.

\paragraph{Simultaneous Approximation for Cutwidth and MLA.} In Section \ref{sec:sim-linear},
we show that the algorithm in Theorem \ref{thrm:CW} simultaneously provides an $ O(\beta(n) \log{n})$-approximation for MLA.
Formally, for an ordering $\pi$ of $V$, let us define $ \mathrm{L}_{\pi}(G)\triangleq \sum _{i\in [n]}|\delta (S^{\pi}_i)|$ to be the linear arrangement cost for $\pi$, and let $ \mathrm{L}(G)\triangleq \min _{\pi} \mathrm{L}_{\pi}(G)$. 

\begin{theorem}\label{thrm:LA}
    The algorithm in Theorem \ref{thrm:CW} finds a linear ordering $\pi$  that simultaneously satisfies:
    
    (i) The cutwidth cost $\mathrm{CW}_{\pi}(G) = O(\beta(n) \log{n})  \cdot \mathrm{CW}(G)$.
    
    (ii) The linear arrangement cost $ \mathrm{L}_{\pi}(G) =   O(\beta(n) \log{n}) \cdot \mathrm{L}(G)$.
\end{theorem}

\paragraph{Vertex Separation Number and Minimum Pathwidth.}
In Section \ref{sec:pathwidth}, we consider a natural vertex variant of cutwidth called the {\em vertex separation} (VS) number.
It was introduced by Lengauer \cite{lengauer81} who used it to define a vertex separator game and relate it to a black-white pebble game in complexity theory. Later, close connections to various other graph parameters were discovered such as the {\em search number} \cite{EHT87}, and  {\em pathwidth} \cite{kinnersley1992vertex}.

\begin{definition}[Vertex Separation Number]
Given a graph $G=(V,E)$ and an ordering $\pi$ of $V$, let
\[ \mathrm{VS}_{\pi}(G) \triangleq \max_{i\in [n]} 
  \big{|}\{ u \in V \,|\, \exists v\in V: (u,v)\in E, \pi(u) \leq i  < \pi(v) \} \big{|}. \]
Then the VS number of $G$ is defined as $\mathrm{VS}(G)\triangleq  \min _{\pi} \mathrm{VS}_{\pi}(G)$.
 \end{definition}
More visually, 
given $\pi$, create an interval $I_u = [\pi(u),\pi(v))$ 
for each vertex $u$,
where $v$ is the furthest neighbor of $u$ to its right ($I_u=\emptyset$ if no such neighbor exists).
Then $\mathrm{VS}_{\pi}(G)$ is the maximum number of intervals crossing any point.

We show the following result.
\begin{theorem}\label{thrm:VS}
There is an efficient $ O(\beta(n) \log{n})$ approximation for vertex separation number.
\end{theorem}
Our algorithm follows the same high-level approach as that for cutwidth, but requires some key modifications to handle vertex cuts instead of edge cuts (see Section \ref{sec:pathwidth}).
Notice that despite their similarity, $\mathrm{VS}(G)$ and $\mathrm{CW}(G)$ can differ substantially. E.g., for a star on $n$ vertices, $\mathrm{VS}(G)=1$ but $ \mathrm{CW}(G)=\Omega(n)$.

\paragraph{Minimum Pathwidth.}
The well-known graph parameter {\em pathwidth}, denoted by $\mathrm{PW}(G)$, was introduced by Robertson and Seymour \cite{ROBERTSON198339}.
Pathwidth is widely used and studied in structural graph theory and fixed parameter tractability and we define it formally in Section \ref{sec:preliminaries}. 

Remarkably, pathwidth has
multiple equivalent definitions, see, { e.g.}, \cite{BODLAENDER19981}.
In particular,
Kinnersley \cite{kinnersley1992vertex} showed that $\mathrm{VS}(G)=\mathrm{PW}(G)$ for all graphs $G$. Moreover, given an ordering $\pi$ with vertex separation number $k$, i.e., $ \mathrm{VS}_{\pi}(G)=k$, there is an efficient algorithm to obtain a path decomposition (see Section \ref{sec:preliminaries} for definition) with pathwidth at most $k$. 

Together with Theorem \ref{thrm:VS}, this directly implies the following.
\begin{theorem}\label{thrm:PW}
There is an $ O(\beta(n) \log{n})$ approximation for \PW.
In particular, for any graph $G$, the algorithm finds a path decomposition with pathwidth $O(\beta(n) \log{n} )\cdot \mathrm{PW}(G)$.
\end{theorem}
The previous approximation for \PW on general graphs was $O(\log^{3/2} n)$ \cite{feige2005improved}.\footnote{\cite{feige2005improved} in fact provide $ O(\log^{1/2}{(\mathrm{CW}(G))}\cdot\log{n})$-approximation for \PW.}
There has also been extensive work on improved approximation bounds for special cases, that we discuss briefly in  Section \ref{sec:RelatedWork}, 
and improving the bound for general graphs is a prominent open problem, see e.g.,~\cite{CDFKMS19}. 

All our results also extend directly to the weighted case. 

\subsection{Overview and Our Approach}\label{sec:overview}
To get some intuition on why min-max objectives are harder than their $\min$-sum counterparts for graph partitioning and layout problems,
let us briefly see why the key techniques for MLA do not work for cutwidth.


\noindent {\bf Cutwidth vs. MLA.}
One way to get an $O(\log{n})$-approximation for MLA is via embedding the metric given by the LP solution (details in Section \ref{sec:preliminaries}) into a distribution over tree metrics 
\cite{FRT04} and finding the best solution.
Such approaches fail for cutwidth as each tree solution may have some large cut (these approaches control each cut in expectation, but we need to bound the expectation of the maximum cut).

For direct approaches based on recursive partitioning, the problem is that some cuts can accumulate too many edges. 
For example, consider Seymour's refined cutting lemma \cite{even2000divide,seymour1995packing} 
which gives a $ O(\log{n}\log{\log{n}})$ approximation for MLA.
In the context of cutwidth, this lemma can return an arbitrary subset $S$ with at most $n/2$ vertices and $|\delta(S)|\leq C \cdot (|S|/n)\cdot \log(n/|S|)\cdot O(\log{\log{n}})$, where $C=\mathrm{CW}(G)$.

\begin{figure}[hbtp!]
  \centering
    \includegraphics[width=0.465\linewidth]{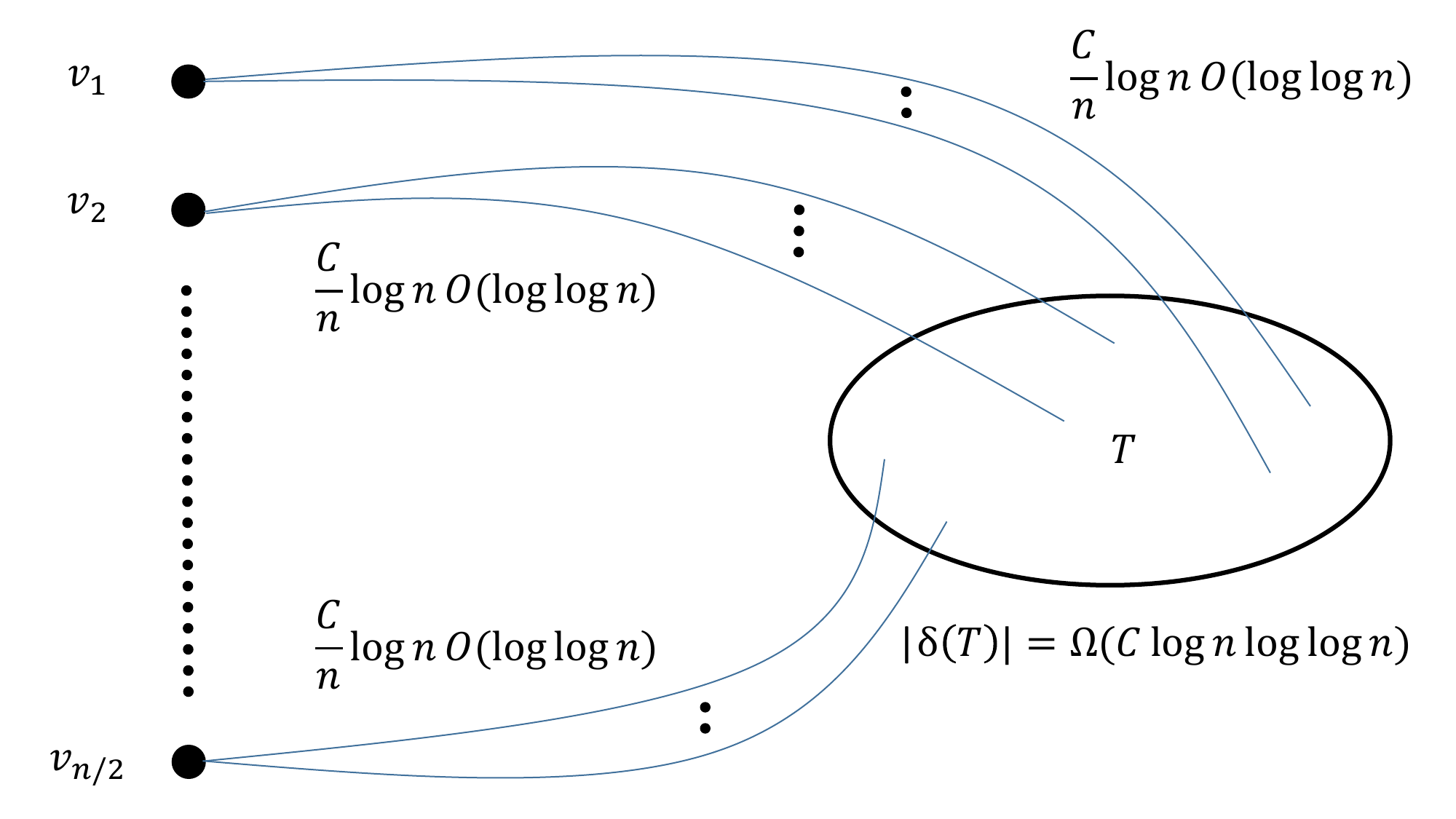}
    \caption{
   Failure of recursive partitioning for cutwidth via Seymour's improved cutting lemma.
    }
    \label{fig:Seymour}
\end{figure}

Suppose applying this lemma repeatedly produces singleton vertex pieces $ \{ v_1\},\ldots,\{ v_{n/2}\}$ for the first $n/2$ steps, so that $ |\delta(\{ v_i\})|=(C/n)\cdot \log{(n)} \cdot O(\log{\log{n}})$ for every $ i$.
However, these $ |\delta (\{ v_1\})|+\ldots +|\delta (\{ v_{n/2}\})| = \Omega (C\log{n}\log{\log{n}} )$ edges could all end up in the remaining piece $T\subset V$ of size $n/2$ (see Figure \ref{fig:Seymour}).
This is problematic 
as repeating this on $T$ recursively will lead to an overall cutwidth of $ \Omega (C\log^2{n}\log{\log{n}} )$.

Similar examples show that other approaches for MLA \cite{rao2005new,charikar2010ℓ,feige2007improved}  also fail for cutwidth.

\paragraph{Our Approach.}
To get around this difficulty, we develop a different decomposition procedure that partitions $G$ into multiple pieces at once (instead of just two above), while ensuring that larger pieces have relatively fewer edges crossing them. This avoids the problem in the example above, as the algorithm recurses on the large pieces. We believe that these ideas could also be useful in other settings (Section \ref{sec:discussion} has some further discussion).

Without going too much into the technical details, we use three key ideas: (i) we partition the graph by choosing balls around a carefully chosen {\em net} of vertices; (ii) we use {\em super-exponential scales} to categorize the vertices and bound the net size; and (iii) we place the pieces in a {\em special order} in the layout.
Perhaps
surprisingly, this allows us to separately control the relative number of edges crossing large and small pieces, while using the standard cutting lemma of Bartal \cite{Bartal96} that is used typically for $\min$-sum objectives.

\subsection{Other Related Work}\label{sec:RelatedWork}
Since its introduction, cutwidth  has found a wide range of applications, {e.g.}, in VLSI design \cite{BL84,makedon1989minimizing}, scheduling \cite{kayaaslan2018scheduling} and protein engineering \cite{BLIN2008618}.
\CW is fixed parameter tractable \cite{thilikos2005cutwidth,giannopoulou2019cutwidth} and admits an exact exponential time algorithm \cite{BFK12}.
It has also been studied for special classes of graphs, {e.g.}, dense graphs \cite{AFK02}, unweighted trees \cite{yannakakis1985polynomial} (weighted trees are NP-hard \cite{MONIEN1988209}), constant treewidth graphs \cite{korach1993tree}, and graphs with bounded vertex cover \cite{cygan2014cutwidth} (see also \cite{ thilikos2005cutwidth, thilikos2005cutwidth2} and the references therein). On the hardness side, assuming the small set expansion hypothesis,
 no constant approximation is possible for cutwidth \cite{WAT14}.

Pathwidth is closely related to treedwidth
and
 has found several applications since its introduction.
 For example, exact algorithms for various NP-hard and $\#$P-complete problems on graphs with constant pathwidth are known 
 (a nice reference is the book by Fomin and Kratsch \cite{FK10}).
 It is also known that the problem of Grundy coloring \cite{BKLM22} is fixed parameter tractable for pathwidth but $ W[1]$-hard for treewidth, thus separating the two parameters. 
 Some practical applications include,
 compiler design \cite{BGT98}, VLSI design \cite{FG94} (see also \cite{Möhring1990} and the references therein), and natural language processing \cite{KORNAI199287}.

There has also been work on computing the pathwidth of a graph. Computing pathwidth exactly  is  known to be fixed parameter tractable \cite{BODLAENDER1996358}, and algorithms  for various special classes of graphs are known (see the survey by Bodlaender \cite{Bod93} and more recently \cite{Groenland21}).

\subsection{Organization}
The rest of the paper is organized as follows. In Section \ref{sec:preliminaries} we give some notation, definitions and some basic concepts we need from previous work.
In Section \ref{sec:cutwidth} we describe our algorithm and analysis for cutwidth, and Section \ref{sec:sim-linear} considers the simultaneous approximation in Theorem \ref{thrm:LA}. In Section \ref{sec:pathwidth} we consider the vertex separation number (equivalently pathwidth).
Finally, in Section \ref{sec:discussion} we describe several open questions and directions for further  exploration that arise from our work.
\section{Notation and Preliminaries}\label{sec:preliminaries}
Throughout this paper we only consider undirected graphs (except in Section \ref{sec:DirectedLAyout} where we discuss directed graphs). For any subset $S \subseteq V$, $G[S]$ denotes the subgraph of $G$ induced on $S$ and $n$ denotes the number of vertices in $G$.

\paragraph{LP Relaxation for Minimum Linear Arrangement.}\label{sec:lp}
Let us recall the standard LP relaxation for MLA, see, {e.g.},~\cite{even2000divide}.
Here, we impose a (semi) metric $d$ on $V$, using the variables $d(u,v)$ for every pair of vertices $ u,v\in V$, that satisfy the spreading constraints \eqref{LP:spread}. 
\begin{align}
(\mathrm{LP})~~~~~~~\min~~~ &  \sum_{(u,v) \in E} d(u,v) \notag \\
\text{s.t.}~~~& d(u,v)+d(v,w) \geq d(u,w) & \forall u,v,w\in V \label{LP:triangle} \\
& \sum_{v\in S} d(u,v) \geq (|S|^2-1)/4  &\forall S \subseteq V, u \in V 
\label{LP:spread} \\
& d(u,v)=d(v,u), \,d(u,u)=0, \, d(u,v)\geq 0  & \forall u,v\in V \label{LP:symmetry}
\end{align}
It is well-known that (LP) can be solved efficiently and is a relaxation for MLA, e.g., \cite{even2000divide}. We sketch the argument below for completeness.

Given  some optimal (integral) linear ordering $\pi ^*$  of $G$,
define $ d(u,v)\triangleq \left| \pi^*(u)-\pi^*(v)\right|$. Clearly, this satisfies the triangle inequalities (\ref{LP:triangle}).
The spreading constraints (\ref{LP:spread}) hold trivially for $|S|=1$. 
Otherwise for $|S|\geq 2$, the left hand side is minimized when $
\lceil|S\setminus \{ u\}|/2\rceil$ vertices of $ S\setminus \{ u\}$ are placed to the left of $u$ and the rest to the right of $u$, which yields the value
$(|S|-1)/2 \cdot ( (|S|-1)/2+1) = \left( |S|^2-1\right)/4$. 
Finally, we note that objective of $ (\mathrm{LP})$ for this solution $d$ is exactly $\mathrm{L}_{\pi^*}(G)$.

For every $x\in V$ and radius $ r\geq 0$, define the ball of radius $r$ centered at $x$ as: 
    \[ B(x,r) \triangleq \{ y\in V : d(x,y) \leq r \}.
    \]
The spreading constraints (\ref{LP:spread}) give the following useful property.
\begin{lemma}\label{lem:size-ball}
For every  $x\in V$ and $ r \geq2$, it holds that $ \left| B(x,r/8)\right|\leq r/2$.  In particular, if $ |V|\geq 2$ then for every $x\in V$: $  \left| B(x,n/8)\right|\leq n/2$. 
\end{lemma}
\begin{proof}
Let $ S=B(x,r/8)$.
If $ |S|=1$ the claim is trivial as $ r\geq 2$.
For $ |S|> 1$, using (\ref{LP:spread}) for $S$ and as $d(x,y) \leq r/8$ for all $y\in S$ and $d(x,x)=0$, we get:
\[ ( |S|^2-1)/4 \leq \sum _{y\in S}d(x,y) \leq \left( |S|-1\right)r/8.\]
Dividing by $|S|-1$ gives $ |S|+1\leq r/2$ and hence $|B(x,r/8)|\leq r/2$.
\end{proof}

We will use $\mathrm{L}^*(G)$ to denote the  optimum  cost of this LP for $G$. Clearly, $\mathrm{L}^*(G) \leq \mathrm{L}(G)$, and  $\mathrm{L}^*(G) = \Omega( \mathrm{L}(G)/\log n)$ by \cite{rao2005new}. 

\paragraph{Lower Bounding Cutwidth.}
To approximate cutwidth, a first question is how to lower bound $\mathrm{CW}(G)$. Clearly, one has the bound\footnote{Indeed, if $\pi$ is some optimum ordering for cutwidth, then $\mathrm{L}(G) \leq \mathrm{L}_\pi(G) \leq n \, \mathrm{CW}_\pi(G) = n \, \mathrm{CW}(G)$.} $\mathrm{CW}(G)\geq \mathrm{L}(G)/n$,  
but this can be arbitrarily weak as one can increase $n$ arbitrarily by adding dummy isolated vertices without affecting $\mathrm{CW}(G)$ and $\mathrm{L}(G)$.


This can be strengthened as follows: 
\[
     \mathrm{CW}(G) \geq \max_{S\subseteq V: S \neq \emptyset} \mathrm{L}(G[S])/|S| , 
 \]
by considering the induced graphs $G[S]$ and noting that 
$ \mathrm{CW}(G)\geq \mathrm{CW}(G[S])$ for any subset $S \subseteq V$ (as any ordering of $V$ also induces a valid ordering of $S$) and $\mathrm{CW}(G[S]) \geq \mathrm{L}(G[S])/|S|$.  

\noindent {\em Remark:} Interestingly, while the lower bound $\mathrm{L}^*(G)$  on $\mathrm{L}(G)$ is easy to compute using the spreading metric LP for MLA, it does not seem to be known how to compute $\max_{S\subseteq V: S \neq \emptyset} \mathrm{L}^*(G[S])/|S|$, even approximately. Interestingly, in Section \ref{sec:LPrelaxations}, we describe a novel LP for cutwidth based on flow metrics \cite{BV04}, which gives a lower bound that is at least  $\max_{S\subseteq V: S \neq \emptyset} \mathrm{L}^*(G[S])/|S|$. 
 However, as our algorithm in Section \ref{sec:cutwidth} does not use this bound  we defer this discussion to Section \ref{sec:LPrelaxations}. 

Instead, our algorithm for cutwidth in Section \ref{sec:cutwidth} uses this lower bound indirectly for specific sets $S$.
  It starts by computing an optimum LP solution for MLA on $G$ and uses the resulting metric $d$ to decompose $V$ into {\em ordered} pieces $S_1,S_2,\ldots,S_k$ (this is our key subroutine). Then, the algorithm recurses on each piece $S_i$ with $|S_i|>1$ (recomputing the LP solution for MLA for each $G[S_i]$) to obtain the final ordering.
Hence, we use the lower bound $\mathrm{L}^*(G[S])/|S|$ only for the sets $S$ produced during the course of the algorithm.

\paragraph{Cutting Lemma for Metric Decomposition.}\label{sec:cutting_lemma}
We use the following standard cutting lemma for metric decomposition based on the well-known exponentially ball growing technique \cite{Bartal96,Bartal98} (see Algorithm \ref{alg:cutting}). For completeness, Algorithm \ref{alg:cutting} and the proof of Lemma \ref{lem:cutting} appear in Appendix \ref{apx:CuttingLemmaProof}.

\begin{lemma}
\label{lem:cutting}
Given an undirected graph $ G=(V,E)$, a set of $T$ terminals $ \{ v_1,\ldots,v_T\}$, a (semi) metric $(V\cup \{ v_1,\ldots,v_T\},d)$, and a radius parameter $ R>0$, 
there is an efficient algorithm that
returns disjoint sets $S_1,\ldots,S_{T}\subseteq V$ satisfying the following properties (with probability $1$):
\begin{enumerate}
\item $ \sum _{t\in [T]}|\delta(S_t)| \leq \frac{8\ln{(2T)}}{R}\sum _{(u,v)\in E}d(u,v)$. \label{decompose1}
\item $ S_t\subseteq B( v_t,2R) \cap V$ for every $ t\in [T]$.\label{decompose2}
\item $ \cup _{t\in [T]} S_t \supseteq \left( \cup _{t\in [T]}B(v_t,R)\right) \cap V$ \label{decompose3}
\end{enumerate}
\end{lemma}

Notice the slight technicality that the terminals $\{ v_1,\ldots,v_T\}$ do not necessarily have to lie in the vertex set $V$. We will need this flexibility when we use this lemma later. 

\paragraph{Pathwidth.} To define the pathwidth of a graph $G$,
we first recall the notion of path decomposition, given by Robertson and Seymour \cite{ROBERTSON198339}.  A path decomposition of $G$ is a sequence $(X_1, X_2, \ldots ,X_r)$ of subsets (bags) $X_i \subseteq V$ for $ i\in [r]$, such that the following three conditions hold:
$(1)$ $\bigcup_{i\in [r]} X_i = V$; $(2)$ for each edge $(u,v) \in E$ there exists some bag $X_i$ containing both $u$ and $v$; and $(3)$ for every indices $i \leq j \leq k \in [r]$, the bags satisfy: $X_i \cap X_k \subseteq X_j$.
The width of a path decomposition $(X_1, \ldots ,X_r)$ is defined as $\max_{i\in [r]}(|X_i|-1)$, and the pathwidth $ \mathrm{PW}(G)$ of a graph $G$ is the minimum width over all path decompositions of $G$.
In the \PW problem, given a graph $G$, the goal is to find $ \mathrm{PW}(G)$.

\section{Approximation for Minimum Cutwidth}
\label{sec:cutwidth}
In this section we consider \CW.
We begin with some definitions and observations in Section \ref{sec:CWsetup}. The Algorithm is described in Section \ref{sec:CWalg} and we prove  Theorem \ref{thrm:CW} in Section \ref{sec:cutwidth-analysis}. 

\subsection{The Algorithm Setup}
\label{sec:CWsetup}
Given the input graph $ G=(V,E)$, where $ |V|=n$, let $d$ be the (semi) metric on $V$ obtained by solving the MLA relaxation  $(\mathrm{LP})$ in Section \ref{sec:preliminaries} and recall that $\mathrm{L}^*(G)$ denotes the value of the solution.

Our overall algorithm is recursive, and given $ G$ as input it performs the following steps:
\begin{enumerate}
\item {\bf Solve LP.} Solve $ (\mathrm{LP})$ for MLA in Section \ref{sec:preliminaries} for $G$, to obtain the (semi) metric $d$;

\item {\bf Decompose $G$.} Given $d$ the solution to $(\mathrm{LP})$ compute an ordered partition $ V_1,\ldots,V_k$ (for some $k$) of the vertices $V$ of $G$, as described in Algorithm \ref{alg:cutwidth} in Section \ref{sec:CWalg};   and

\item {\bf Recurse.} For every $ r\in [k]$ with $|V_r|>1$, recursively execute the algorithm on the subgraph $ G[V_r]$ induced by $ V_r$ to obtain the ordering within $V_r$.
\end{enumerate}
This produces the final ordering of the vertices.

\noindent{\em Remark:} Notice that in Step (3), when recursively executing on subgraph $ G[V_r]$, the algorithm re-solves $(\mathrm{LP})$ for $G[V_r] $ (and does not use the original $ (\mathrm{LP})$ solution for $G$). 

Clearly,
the heart of the algorithm is  the decomposition procedure in Step $(2)$.

\subsection{Radius and Size Scales}
\label{sec:scales}
Before describing the decomposition procedure, we define two key notions of {\em radius scale} and {\em size class}, and note some simple properties they satisfy.

Define the sequence $ \Delta _i\triangleq n/8^i$ for $ i=1,2,\ldots$.
For each $v$, we have $B(v,\Delta_1)\supseteq B(v,\Delta _2)\supseteq \ldots$ as the $\Delta_i$ are decreasing,
and thus the sequence $ \{n/|B(v,\Delta_i)|\}_{i=1,2\ldots}$ is non-decreasing.

For each $v$, we will be interested in the smallest index $i$ where the increase from $ n/|B(v,\Delta _{i-1})|$ to $n/|B(V,\Delta _i)|$ is not too large.
To quantify this, let $ \gamma \triangleq \gamma(n)\gg 1$ be a parameter that we set to $\exp(O((\log \log n)^{1/2}))$ in hindsight\footnote{We assume that $\log$ refers to the logarithm of base $2$.}, but for now we only assume that  $\gamma \geq 2$.

\begin{definition}[Radius Scale]\label{def:radius-scale}
    For every vertex $v\in V$, the radius scale $i_v$ of $v$ is the smallest index $i$, where $ i\geq 2$, such that:
    \begin{equation}
        \log \left(\frac{n}{|B(v,\Delta_i)|}\right) \leq \gamma \cdot \log \left(\frac{n}{|B(v,\Delta_{i-1})|} \right).\label{eq:radius-scale}
    \end{equation}
\end{definition}
The balls $B(v,\Delta_{i_{v}})$ and $B(v,\Delta_{i_v-1})$ for each $v$ will play an important role.
We will denote $B_v \triangleq B(v,\Delta_{i_{v}})$ and call $B_v$ the ball of $v$. We will sometimes use $2B_v$ to denote the ball $B(v,2\Delta_{i_{v}})$. 

Let us recursively define a sequence $\{ n_k\}_{k=1}^{\infty}$ related to \eqref{eq:radius-scale}
as follows:
\begin{align}
   \begin{cases}
    n_1= \frac{n}{2} & \\
    \log{\left(\frac{n}{n_k}\right)} = \gamma \cdot \log{\left(\frac{n}{n_{k-1}}\right)} & k>1.
   \end{cases} \label{n_sequence}
\end{align}
In other words, for every $k\geq 1$, we have $ \log{(n/n_k)}=\gamma ^{k-1}$, or equivalently, $ n_k=n/2^{\gamma^{k-1}}$.
Also, note that
$n_{\ell}\leq 1$
for $ \ell \triangleq \lceil \log_\gamma \log n \rceil + 1$. This $\ell$ will be the number of {\em scales} in our algorithm.

Notice that $ n_1,\ldots,n_{\ell}$ would be the sizes of the balls $ B(v,\Delta_1),\ldots, B(v,\Delta _{\ell})$ if $|B(v,\Delta _1)|=n/2$, and
 (\ref{eq:radius-scale}) was tight for every $ i=2,\ldots,\ell$.

\begin{definition} [Size Class]\label{def:size-class}
We say a vertex $v$ has size class $j_v$ if $|B_v|  \in (n_{j_v},n_{j_v-1}]$.
\end{definition}

The following Lemma gives some simple but useful observations, and is proved in Appendix \ref{app:NewBallBound}.
\begin{lemma}
\label{fact:NewBallBound}
    For every $v$, the quantities $i_v,j_v \in [2,\ell+1]$. Moreover, $|B(v,\Delta _{i_v-1})| \leq n_{j_v-2}$, where we use the convention $n_0=n/2$ to handle $j_v=2$. In particular, $ |B(v,\Delta_{i_v-1})|\leq n/2$ for all $v$.
\end{lemma}

\subsection{The Decomposition Procedure}\label{sec:CWalg}
We now describe the procedure to decompose $V$ into pieces $V_1.\ldots,V_k$ and how to order these pieces. This consists of the following three phases (also depicted in Algorithm \ref{alg:cutwidth}):
\begin{enumerate}
\item{\bf Compute $i_v,j_v$.}
Compute the radius scale $i_v$ and size class  $j_v$ for each vertex $v$.
Let $C_i$ be the set of  vertices with radius scale $i$. Then 
$\{C_2,\ldots,C_{\ell+1}\}$ is a partition of $V$.

\item
{\bf Packing Balls.} For each radius scale $i$, compute a maximal
disjoint packing of balls $B_v$ for $v \in C_i$.
Let $N_i$ denote the ``net" consisting of the centers $v$ of the balls in this packing.

Next, we further partition the vertices in $N_i$, according to their size class: 
partition $N_i$ further into $ \{ N_{i,j}\}_{j=2}^{\ell +1}$, where 
$N_{i,j}$ consists of vertices in $N_i$ with size class $j$.

\item {\bf Apply Lemma \ref{lem:cutting}.} 
 Iterate over the size classes $j=2,3,\ldots,\ell+1$ in increasing order, and for each $j$, iterate over the radius scales $i=2,3,\ldots,\ell+1$ in increasing order.\footnote{The order over $i$ will not matter in the  analysis, but we fix it here for concreteness. However, the order over $j$ will be very crucial for our algorithm and analysis.}

In each iteration $(j,i)$: Apply the algorithm of Lemma \ref{lem:cutting} on the remaining graph, with terminals $ N_{i,j}$ and parameter $R = 2 \Delta_{i}$. Return the resulting pieces $S_v$, and remove them from the graph.
\end{enumerate}
\RestyleAlgo{ruled}
\begin{algorithm}[H]
\caption{Partition and Arrange $ (G=(V,E),d)$}\label{alg:cutwidth}
\texttt{/* Phase $1$ - preprocessing */}\\
$\forall v\in V$: Compute $i_v$ radius scale and $j_v$ size class of $v$.\\
$\forall i=2,\ldots,\ell+1$: $ C_i\leftarrow \{ v\in V:i_v=i\}$.\\
\texttt{/* Phase $2$ - nets computation */}\\
\For{$ i=2,\ldots,\ell+1$}{
$ N_i\leftarrow \emptyset$.\\
\While{$ \exists v\in C_i$ s.t. $ {{B_v}}\cap (\cup _{u\in N_i} {{B_u}})=\emptyset$}{
$ N_i\leftarrow N_i\cup \{ v\}$.
}
$\forall j=2,\ldots,\ell+1$: $ {{N_{i,j}}}\leftarrow N_i\cap \{ v\in V:j_v=j\}$.
}
\texttt{/* Phase $3$ - decomposing and ordering */}\\
Set  $V' \leftarrow V$.\\
 \For{$ j=2,\ldots,\ell+1$}{
 \For{$ i=2,\ldots,\ell+1$}{
 Apply Lemma \ref{lem:cutting} with parameters $(G[V'],{{N_{i,j}}},d,2\Delta _i)$ to obtain the pieces $\{ S_v\}_{v\in {{N_{i,j}} }}$.\\
$V' \leftarrow V' \setminus (\cup_{v\in {{N_{ij}} }} S_v)$
 }
}
Arrange the pieces in the order they were produced.
\end{algorithm}

\subsection{Analysis}\label{sec:cutwidth-analysis}
We now show that the cutwidth produced by the algorithm is at most $\beta(n) \cdot \mathsf{CW}(G) \cdot\log n$.
Consider the pieces produced by the algorithm (arranged in the order they were constructed). Each such piece  is of the form $S_v$, for some $ v\in N_{i,j}$. 
Recall that $N_i = N_{i,2}\cup \ldots \cup N_{i,\ell+1}$ and let us denote $N  \triangleq N_2\cup \ldots \cup N_{\ell+1}$. For $u,v \in N$, we write $S_u \prec S_v$ if the piece $S_u$ was produced before $S_v$, and write  $S_u\preceq S_v$ if $ S_u\prec S_v$ or $ u=v$.\footnote{For fixed $i$ and $j$ the order of pieces in $ \{ S_v\}_{v\in N_{i,j}}$ does not matter for our analysis.
However, for concreteness we order the pieces in $ \{ S_v\}_{v\in N_{i,j}}$ according to the order the algorithm of Lemma \ref{lem:cutting} constructs them.}

The result will follow directly from the next two lemmas.

\begin{lemma}
\label{lem:bound-size}
The pieces $\{S_v\}_{v \in N}$ partition $V$.
Moreover, $|S_v| \leq n_{j_v-2} \leq  n/2$
for all  $v \in N$.
\end{lemma}
In particular, our algorithm finds a non-trivial partition of $V$ and thus eventually terminates.
\begin{lemma}
    \label{lem:pass-over}
For every $v\in N$, the total number of edges cut by pieces produced until $S_v$ is created  (and including $S_v$ itself) satisfies:
\[ \sum_{w : S_w \preceq S_v} |\delta(S_w)| \leq  8^{\ell+4} \gamma^{2} \cdot \log\left(n/|S_v| \right) \cdot \mathrm{CW}(G). \]
\end{lemma}
Before proving Lemmas \ref{lem:bound-size} and \ref{lem:pass-over}, we first show they directly give Theorem \ref{thrm:CW}.

\paragraph{Bounding the Overall Cutwidth.}
Let $\mathsf{Alg}(G)$ denote cutwidth of $G$ for the final ordering $\pi$ produced by the recursive algorithm. 
Consider some position $r \in [n]$ in this ordering.
We bound the number of edges crossing the cut $ S_r^{\pi}$ (recall that $ S_r^{\pi}=\{ y\in V:\pi (y)\leq r\}$), i.e., $ |\delta (S_r^{\pi})|$. 
Let $S_v$ be the piece produced by Algorithm \ref{alg:cutwidth} such that some vertex from $S_v$ is placed at position $r$.

\begin{figure}[hbtp!]
  \centering
    \includegraphics[width=\linewidth]{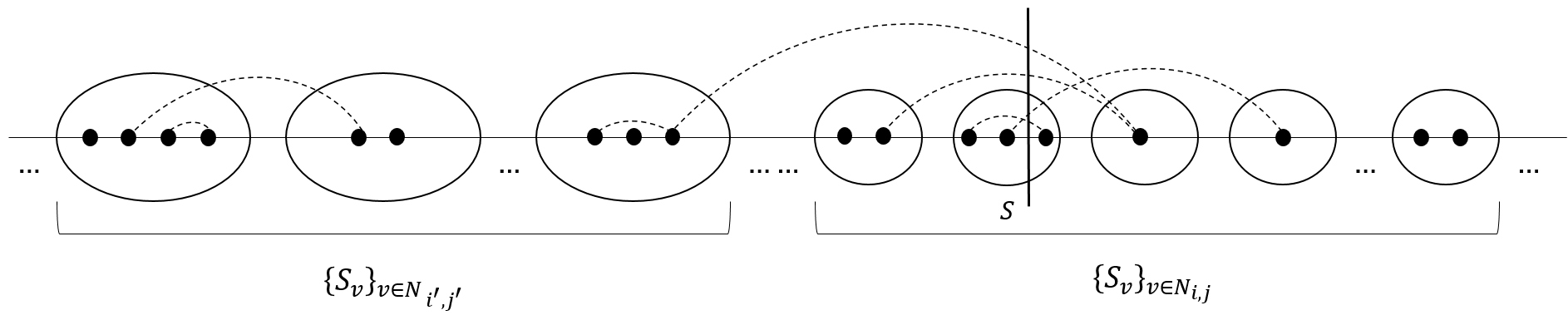}
    \caption{
   Ordering of the pieces $S_v$ by Algorithm \ref{alg:cutwidth}. 
    The edges passing over a piece $S\in \{S_v\}_{v\in N_{i,j}}$ must lie in some $ \delta(S_v)$, for some $ v\in N_{i',j'}$ and $j' \leq j$, or in $ G[S]$.
    }
    \label{fig:cutwidth}
\end{figure}

Then the number of edges crossing the cut $S_r^{\pi}$ is at most the sum of the following two quantities (see also Figure \ref{fig:cutwidth}): 

(i)
$\mathsf{Alg}(G[S_v])$ due to the edges produced by the algorithm within $S_v$ recursively; and 

(ii) number of edge between all distinct pieces $S_w$ and $S_{w'}$ where $w \preceq v$ and $v \prec w'$, which is at most $\sum_{w: S_w \preceq S_v} |\delta(S_v)|$.

Together with Lemma \ref{lem:pass-over}, we obtain the following bound on the cutwidth of the algorithm: 
\begin{equation}
    \label{eq:basic-recursion}
    \mathsf{Alg}(G) \leq   \max_{v \in N} \left( 8^{\ell+4} \gamma^2\cdot  \mathrm{CW}(G)  \cdot \log (n/|S_v|)) +   \mathsf{Alg}(G[S_v]) \right) .
\end{equation}

\paragraph{Optimizing $\beta(n)$ and Choosing $\gamma$.} Let us denote $\beta(n)\triangleq  8^{\ell+4}\gamma^2$ (the factor appearing in \eqref{eq:basic-recursion}), and recall from Section \ref{sec:CWsetup} that $ \ell = \ell(n) =\lceil \log_\gamma \log n \rceil + 1 = O(\log \log n)/\log \gamma$.
Then,
\[  \log \beta(n) = O(\ell + \log \gamma) = O\left(\log \log n/\log \gamma + \log \gamma \right),\]
and choosing $\log \gamma = \sqrt{\log \log n}$ gives $
\beta(n) = \exp(O(\sqrt{\log \log n}))$.

\paragraph{Solving the Recursion.} We now prove by induction on the number of vertices that $ \mathsf{Alg}(G)\leq \beta(n) \cdot \mathrm{CW}(G)\cdot \log{n}$.
Assuming inductively that $ \mathsf{Alg}(G[S_v]) \leq \beta(|S_v|) \cdot \mathrm{CW}(G[S_v]) \cdot \log |S_v|$ for every $ G[S_v]$ where $v\in N$,
\eqref{eq:basic-recursion}  gives that,
\begin{align*}
    \label{eq:basic-recursion}
    \mathsf{Alg}(G) & \leq   \max_{v \in N}  \Big( \beta(n) \cdot  \mathrm{CW}(G)  \cdot \log (n/|S_v|)) +  \underbrace{\beta(|S_v|)}_{\leq \beta(n)} \cdot \underbrace{\mathrm{CW}(G[S_v])}_{\leq\mathrm{CW}(G)} \cdot \log |S_v|  \Big) \\ 
    & \leq   \beta(n) \cdot  \mathrm{CW}(G) \cdot \max_{v \in N} (\log (n/|S_v|)+  \log |S_v|) \\
     & = \beta(n) \cdot  \mathrm{CW}(G)  \cdot \log n .
\end{align*}
This concludes the proof of Theorem \ref{thrm:CW}.

We now prove Lemmas \ref{lem:bound-size} and \ref{lem:pass-over}.


\begin{proof}[Proof of Lemma \ref{lem:bound-size}]
   Clearly, by design of Algorithm \ref{alg:cutwidth} and Lemma \ref{lem:cutting} the pieces $S_v$ produced are always disjoint (once pieces are produced by Lemma \ref{lem:cutting}, Algorithm \ref{alg:cutwidth} removes them from the graph before producing the next pieces). Thus, for the first part it suffices to show that they cover $V$.

We first show  that $\cup_{v\in N}2B_v = V$.
This follows from the classical packing-covering argument.
Suppose there is some vertex $ w \notin
\cup_{v\in N}2B_v$, and let $i=i_w$ denote its radius scale.
Then, at the very least, $B_w$ is disjoint from $B_v$ for every vertex $v \in C_i$ with radius scale $i$. But as the nets $N_i$ are created separately for each scale, $w$ should have been added to 
$N_i$, contradicting its maximality.
    
    
    Now, consider some piece $S_v$, where $ v\in N_{i,j}$.
Requirement (\ref{decompose3}) of Lemma \ref{lem:cutting} implies that $ \cup _{v\in N_{i,j}}S_v \supseteq ( \cup _{v\in N_{i,j}}2B_v) \cap V$, where $V$ is the set of vertices that were not covered so far by Algorithm \ref{alg:cutwidth}.
The reason for this is that $ 2B_v=B(v,2\Delta_{i_v})=B(v,R)$ (recall that $ R=2\Delta _i$).
Thus, $\cup _{v\in N_{i,j}} S_v$ will contain all the vertices  in $( \cup _{v\in N_{i,j}}2B_v)$, except possibly those that are already covered earlier by pieces $S_w$, where $ w\in N_{i',j'}$ with $j'<j$ (or $j'=j$ and $ i'<i$). 
This gives that $\cup_{v \in N} S_v \supseteq \cup_{v \in N} 2B_v$, where $ \cup_{v \in N} 2B_v=V$ by the argument above.

The second part follows by noting that requirement (\ref{decompose2}) of Lemma \ref{lem:cutting} implies $ S_v\subseteq B(v,2R)\cap V=B(v,4\Delta_i)\cap V$ (since $ R=2\Delta_i$).
Thus, by Lemma \ref{fact:NewBallBound},   $|S_v| \leq n_{j_v-2} \leq n/2$.
\end{proof}

\begin{proof}[Proof of Lemma \ref{lem:pass-over}]
Fix some $v \in N$ and suppose that $v\in N_{i,j}$. As the nets $N_{i,j}$ are considered in an increasing order of $j$ in Phase 3 of Algorithm \ref{alg:cutwidth}, we have that $j_w \leq j$ for any piece $S_w \preceq S_v$ that is constructed  prior to $S_v$ (including $S_v$ itself).
Thus, we have the following bound: 
\begin{equation}
     \sum_{w : S_w \preceq S_v} |\delta(S_w)| \leq \sum_{j'=2}^{j} \sum_{i'=2}^{\ell+1} \sum_{w \in N_{i',j'}} |\delta(S_w)|.
     \label{eq:cross-edges-bound}
\end{equation}
We will show that each term on the right of (\ref{eq:cross-edges-bound}), for every $i' \in [2,\ell+1]$ and $ j'\in [2,j]$, can be bounded as follows:
\begin{equation}
    \sum_{w \in N_{i',j'}} |\delta(S_w)| \leq  8^{i'+2} \gamma^{j'-1} \cdot \mathrm{CW}(G).
        \label{eq:cross-edges-bound-ij}
\end{equation}

The pieces $\{S_w\} _{w\in N_{i',j'}}$ are obtained by applying Lemma \ref{lem:cutting} with terminals $ N_{i',j'}$ and $ R=2\Delta_{i'}$ (to the graph that remains at iteration $(j',i')$ in Algorithm \ref{alg:cutwidth}).
Therefore, by requirement (\ref{decompose1}) of Lemma \ref{lem:cutting}:
\begin{align}
 \sum_{w\in N_{i',j'}}|\delta (S_w)| \leq \frac{8\ln{(2|N_{i',j'}|)}}{2\Delta _{i'}}\cdot \mathrm{L}^*(G),\label{eq:crossHij} 
\end{align}
where we trivially bound the total edge lengths $d(u,v)$ in the remaining graph by $\mathrm{L}^*(G)$.

We now claim that $|N_{i',j'}| \leq n/n_{j'}=2^{\gamma^{j'-1}}$.
Indeed, by the definition of $N_{i'}$, the balls $ \{ B_v\}_{v\in N_{i'}}$ are  disjoint, and as $ N_{i',j'}\subseteq N_{i'}$ the balls $\{ B_v\}_{v\in N_{i',j'}}$ are also disjoint. Now as each $v \in N_{i',j'}$ has size class $j'$,  $|B_v|\geq n_{j'}$ (by Definition \ref{def:size-class}) and thus $ |N_{i',j'}|\leq n/n_{j'} $.

Plugging this bound for $|N_{i',j'}|$ in \eqref{eq:crossHij} and as $ \Delta _{i'}=n/8^{i'}$, we obtain:
\begin{align}
\frac{8\ln{(2|N_{i',j'}|)}}{2\Delta _{i'}} \cdot \mathrm{L}^*(G) \leq 8\cdot 8^{i'} \cdot (\gamma ^{j'-1}+1)\cdot \frac{\mathrm{L}^*(G)}{n} \leq 
 8^{i'+2}\gamma ^{j'-1}\cdot \mathrm{CW}(G),\label{eq:crossHij2} \end{align}
 where the last inequality uses that $\mathrm{L}^*(G)/n$ is lower bound on $\mathrm{CW}(G)$. This proves \eqref{eq:cross-edges-bound-ij}.
 
Summing \eqref{eq:cross-edges-bound-ij} over $i' \in [2,\ell+1]$ and $j'\in [2,j]$ (and using that $1+ \gamma + \ldots + \gamma^{j-1} \leq 2 \gamma^{j-1}$) gives:
\[  \sum_{j'=2}^{j} \sum_{i'=2}^{\ell+1} \sum_{w \in N_{i',j'}} |\delta(S_w)| \leq  8^{\ell+4} \gamma^{j-1} \cdot \mathrm{CW}(G) = 8^{\ell+4} \gamma^2\cdot  \mathrm{CW}(G)\cdot \underbrace{\gamma^{j-3}}_{\leq\log (n/n_{j-2})}  \leq  8^{\ell+4} \gamma^2\cdot \mathrm{CW}(G) \cdot\log (n/|S_v|),\]
where the last inequality uses that $|S_v| \leq n_{j_v-2}$ by Lemma \ref{lem:bound-size}. This gives the claimed result.
  \end{proof}

\section{Simultaneous Approximation for Cutwidth and MLA}\label{sec:sim-linear}
We now show that the algorithm for cutwidth described in Section \ref{sec:CWalg} also simultaneously approximates  MLA to within the same approximation factor of $O(\beta(n) \cdot \log n)$.


Let us start with some notations that will be useful.
Let $\Vol(G) =\sum_{(u,v)\in E} d(u,v)$ denote the volume of $G$ given by the solution to $(\mathrm{LP})$, and recall that $\Vol(G)\leq \mathrm{L}(G)$. 
We are now ready to prove Theorem \ref{thrm:LA}.
\begin{proof}[Proof of Theorem \ref{thrm:LA}]
Let us denote $L\triangleq  \Vol(G)$, and let $L_j$ denote the volume of edges removed from $G$ when the cutting procedure, i.e., Lemma \ref{lem:cutting}, at size scale $j$ is executed by Algorithm \ref{alg:cutwidth}.

As $ j\in [2,\ell+1]$, this gives $L = L_2 + \dots + L_{\ell+1}$. 
Let $R_j \triangleq  L - L_2 - \ldots - L_{j-1} = L_{j}+\ldots+L_{\ell+1}$ denote the volume in $G$ immediately
before the cutting procedure at size scale $j$ 
is applied.
By the proof of Lemma \ref{lem:pass-over} and inequality (\ref{eq:crossHij}), the number of edges removed by the cutting procedure at size scale $j$ is at most:
\begin{align}
    \sum _{i=2}^{\ell+1}\sum _{w\in N_{i.j}}|\delta (S_w)|\leq \sum _{i=2}^{\ell+1}\frac{8\ln{(2|N_{i,j}|)}}{2\Delta _i}\cdot R_j,\label{eq:pass-overLA} 
\end{align}
where $R_j$ is used instead of $\mathrm{L}^*(G)$ (as mentioned in the proof of Lemma \ref{lem:pass-over} $ \mathrm{L}^*(G)$ is an upper bound on the sum of lengths of edges that remain in the graph, which in our case is exactly $R_j$).

By inequality (\ref{eq:crossHij2}), the right hand side of (\ref{eq:pass-overLA}) is upper bounded by:
\begin{align}
   \sum _{i=2}^{\ell+1}\frac{8\ln{(2|N_{i,j}|)}}{2\Delta _i}\cdot R_j\leq \gamma ^{j-1}\cdot \frac{R_j}{n}\cdot \sum _{i=2}^{\ell+1}8^{i+2}\leq 2\cdot 8^{\ell+3}\gamma ^{j-1}\cdot \frac{R_j}{n}. \label{eq:pass-overLA2}
\end{align}

Now, the linear arrangement cost of the final ordering for $G$ consists of two parts: (i) arrangement cost of the edges removed when applying Algorithm \ref{alg:cutwidth}; and
(ii) the cost of the linear arrangement obtained by the algorithm recursively on each piece $S_v$, $v\in N$. 

As each removed edge has stretch at most $n$ in the final ordering, the total contribution of (i) above is at most:
\begin{align}
    n\sum _{j=2}^{\ell+1}2\cdot 8^{\ell+3}\gamma ^{j-1}\cdot \frac{R_j}{n} &=2\cdot 8^{\ell+3}\sum _{j=2}^{\ell+1}\gamma^{j-1}(L_j+\ldots+L_{\ell+1})\nonumber \\
    & =2\cdot 8^{\ell+3} \sum _{j=2}^{\ell+1}(\gamma+\gamma^2+\ldots+\gamma^{j-1}) L_j\nonumber \tag{collecting coefficients for each $L_j$}\\
    &\leq 8^{\ell+4}\sum _{j=2}^{\ell+1}\gamma^{j-1}\cdot L_j.  \tag{as $\gamma +\gamma^2+\ldots +\gamma ^{j-1}\leq 2\cdot \gamma^{j-1}$} 
\end{align}

Let $\mathsf{Alg}(L,n)$ be the value of the output of the algorithm on any graph with $n$ vertices and a given a feasible solution $d$ to $(\mathrm{LP})$ with volume $L$. 
Thus, we can bound $\mathsf{Alg}(L,n)$ as follows:
\begin{equation}
    \label{eq:rec-simultaneous}
     \mathsf{Alg}(L,n) \leq \sum_{j=2}^{\ell+1}\left( \sum_{v\in N_j} \mathsf{Alg}(\Vol(S_v),|S_v|) +8^{\ell+4}\cdot  L_j \cdot  \gamma^{j-1} \right). 
\end{equation}
In the above, $ \Vol(S_v)$ denotes the volume of an optimal solution to $ (\mathrm{LP})$ once re-solved for $ G[S_v]$.

We will show by induction that $\mathsf{Alg}(L,n) \leq L \cdot \beta(n)\cdot \log n $, where $\beta(n)=8^{\ell+4}\gamma^2$.
As \eqref{eq:rec-simultaneous} holds whenever $n\geq 2$ (recall the stopping condition is $ n=1$) it suffices to consider the base case $n=1$, where the claim trivially holds. 
We now show the inductive step.

As each $|S_v| \leq n_{j-2}$ for every $v\in N_j$ (Lemma \ref{fact:NewBallBound} and as $ S_v\subseteq B(v,\Delta _{i_v-1})$), applying the inductive hypothesis to each piece $S_v$, where $v\in N_j$, and using that
$ \sum _{v\in N_j}\Vol (S_v) \leq L_j$ (recall that $L_j$ is the total volume removed by the cutting procedure at size class $j$), we get:
\[    \sum_{v\in N_j} \mathsf{Alg}(\Vol(S_v),|S_v|)  \leq  \sum_{v\in N_j} \underbrace{\beta(|S_v|)}_{\leq \beta(n)} \cdot \Vol(S_v) \cdot \underbrace{\log |S_v|}_{\leq \log (n_{j-2})} 
    \leq L_j \cdot \beta(n)\cdot \log (n_{j-2}).\]
Plugging this in \eqref{eq:rec-simultaneous} and using that $8^{\ell+4} \cdot \gamma^{j-1} = 8^{\ell+4} \cdot  \gamma^2 \cdot \gamma^{j-3} \leq \beta(n) \cdot \log (n/n_{j-2})$ (by the definition of $\beta(n)$ and $n_{j-2}$), we get:
 \begin{align*}
   \mathsf{Alg}(L,n)  &\leq \sum_{j=2}^{\ell+1} L_j \cdot \beta(n)\cdot \log (n_{j-2}) + 8^{\ell+4}\sum_{j=2}^{\ell+1} L_j  \cdot \gamma^{j-1}  \\
    &\leq \beta (n) \sum_{j=2}^{\ell+1} L_j \left( \log(n_{j-2}) + \log(n/n_{j-2})\right)\\
    &=\beta (n)\cdot \log{n}\sum_{j=2}^{\ell+1} L_j = L\cdot \beta(n) \cdot\log n. \qedhere
\end{align*}
\end{proof}

\section{Approximation for Pathwidth}\label{sec:pathwidth}
In this section we present an $O(\beta(n) \log n)$ approximation for pathwidth.

\paragraph{LP relaxation for Vertex Separation Number.}
We start with a natural LP relaxation for $\mathrm{VS}(G)$. 
Similarly to $ (\mathrm{LP})$ and $\mathrm{CW}(G)$, the relaxation finds a (semi) metric $d$ over $V$ that satisfies the same spreading constraints (constraints (\ref{LPVS:triangle}), (\ref{LPVS:spreading}) and (\ref{LPVS:standard})).
However, we additionally have variables $x_u$, for every vertex $u\in V$.
Intuitively, every $x_u$ should be viewed as the length of the interval $I_u$ in the visual interpretation mentioned in Section \ref{sec:results} since these variables satisfy constraint (\ref{LPVS:edge}). 
\begin{align}
(\mathrm{LP}_{\mathrm{VS}})~~~~~~~\min~~~ &  \sum_{u \in V} x_u  \nonumber\\
\text{s.t.}~~~& x_u+ x_v \geq d(u,v) & \forall (u,v)\in E \label{LPVS:edge}\\
 &d(u,v) + d(v,w) \geq d(u,w) &\forall u,v,w \in V  \label{LPVS:triangle}\\
 &\sum_{v\in S} d(u,v) \geq \frac{1}{4}(|S|^2-1)  & \forall S \subseteq V, \, \forall u \in V \label{LPVS:spreading}\\
 &d(u,v) = d(v,u), d(u,u)=0, d(u,v)\geq 0 &\forall u,v \in V \label{LPVS:standard}
\end{align}
The following lemma proves that $(\mathrm{LP}_{\mathrm{VS}})$ is indeed a relaxation for $ \mathrm{VS}(G)$.
\begin{lemma}\label{lem:pathwidth-relaxation}
For every graph $ G=(V,E)$ on $n$ vertices, let $P^*$ be the value of an optimal solution to $ (\mathrm{LP}_{\mathrm{VS}})$. Then, $ P^*/n\leq \mathrm{VS}(G)$.
\end{lemma}
\begin{proof}
   Let $\pi :V \rightarrow \{1, \ldots, |V|\}$ be an ordering achieving the optimal vertex separation number.
   As in $ (\mathrm{LP})$ and $\mathrm{CW}(G)$, $\pi$ naturally defines a (semi) metric $d(u,v) \triangleq |\pi(u)-\pi(v)|$ that satisfies the spreading constraints.
   Thus, constraints (\ref{LPVS:triangle}), (\ref{LPVS:spreading}), and (\ref{LPVS:standard}) are satisfied.  
    For every vertex $u\in V$, define $x_u\triangleq  \max_{v\in N(u)} \{\pi(v)-\pi(u),0\}$, where $N(u)$ is the collection of $u$'s neighbors.
    We note that $x_u$ equals exactly the length of $ I_u$ that $\pi$ defines.
    Clearly, constraints (\ref{LPVS:edge}) are satisfied with this choice of $x_u$ values.
Finally, we note that the total length of the intervals equals $ P^*$.
Thus, averaging over the $n$ cuts $\pi$ defines: $S^{\pi}_1,\ldots,S^{\pi}_n$ (recall that $ S^{\pi}_i=\{ u\in V:\pi(u)\leq i\}$) proves that $ P^*/n\leq \mathrm{VS}(G)$.
\end{proof}

Similarly to  $(\mathrm{LP})$ and $ \mathrm{CW}(G)$, the lower bound of $P^*/n $ can be arbitrarily weak, e.g., one can add dummy isolated vertices.
However, as our algorithm is recursive we will solve $(\mathrm{LP}_{\mathrm{VS}})$ on multiple subgraphs of $G$, circumventing this difficulty.

\subsection{The Algorithm}
Our  algorithm for pathwidth is similar to that for cutwidth and it follows the same structure as Algorithm \ref{alg:cutwidth}.
The key differences are in Phase $3$ of the algorithm, and how the cutting procedure in Lemma \ref{lem:cutting} operates, as for pathwidth we need to consider vertices, whereas for cutwidth we considered edges.

\paragraph{Overview.} As in Algorithm \ref{alg:cutwidth} for cutwidth, our algorithm partitions the vertex set $V$ into pieces $V_1, \ldots ,V_k$ (for some $k$) and orders the pieces suitably, say as follows: $V_1,V_2,\ldots,V_k$.
Then, we recurse on each of the pieces $V_r$ to obtain the final ordering.

For $S \subseteq V$, let $\mathsf{Alg}(G[S])$ denote the vertex separator number of the ordering obtained by the algorithm on the subgraph $G[S]$ induced by $S$. 
Also, for every $r\in [k]$ let $X_r$ denote the set of vertices belonging to $V_1 \cup \dots \cup V_r$ that have at least one neighbor in $V_{r+1} \cup \dots \cup V_k$. 
Then, a crucial observation is that we get the following recurrence for $\mathsf{Alg}(G)$ that bounds the value of the output of the algorithm:
\begin{equation}
    \label{eq:vs-recursion}
    \mathsf{Alg}(G) \leq   \max_{r\in[k]} \left( |X_r| +     \Alg(G[V_r]) \right).
\end{equation}

Similarly to the analysis of Algorithm \ref{alg:cutwidth},
the bulk of the effort will be to prove that our algorithm satisfies the following guarantee for every  $r \in [k]$ (with $ |V_r|>1$): 
\begin{align}
|X_r| \leq \frac{P^*}{n} \cdot \beta(n) \cdot \log\left(\frac{n}{|V_r|}\right) \leq \mathrm{VS}(G) \cdot\beta(n) \cdot\log\left(\frac{n}{|V_r|}\right), \label{eq:vs-bound}
\end{align}
where $\beta(n) = \exp(O(\sqrt{\log \log n}))= \log^{o(1)}{(n)}$ as before.
Plugging the bound (\ref{eq:vs-bound}) in the recursive relation \eqref{eq:vs-recursion} results in the desired upper bound of $ \mathsf{Alg}(G)\leq \beta(n)\cdot \mathrm{VS}(G)\cdot \log{n}$.

\paragraph{Algorithm Description.}
First, let us describe the updated metric decomposition algorithm that takes the vertices into consideration, which appears in Algorithm \ref{alg:cutting-vertex}.
Intuitively, one can consider a fixed imaginary ball of radius $x_u$ around every vertex $u\in V$.
Then, as the algorithm progresses, if the ball of radius $R(1+\rho_t) $ around terminal $v_t$ (ball $ B_t$ as in Algorithm \ref{alg:cutting}), cuts this fixed imaginary ball around a vertex $u$, then $u$ is removed and placed in $D_t$.
If this happens, i.e., $R(1+\rho_t)\in [d(v_t,u)-x_u,d(v_t,u)+x_u] $, we say that $u$ is cut by $B_t$.
At the end, $D$ contains all vertices $u\in V$ that were cut in this manner.
Also, we make sure that all the pieces $S_t$ are disjoint to the vertices that are cut and belong to $D$.
\RestyleAlgo{ruled}
\begin{algorithm}[hbt!]
\caption{Metric Vertex Decomposition $ (G=(V,E),\{ v_1,\ldots,v_{T}\},d,\{ x_u\}_{u\in V},R)$}\label{alg:cutting-vertex}
$\lambda \leftarrow \ln{(2T)}$.\\
Sample $ \rho_1,\ldots,\rho_{T}$ i.i.d distributed  $ \text{exp}(\lambda)$ repeatedly, until $ \rho _t\leq 1$ $ \forall t\in [T]$.\\
$ D\leftarrow \emptyset$.\\
\For{$ t=1,\ldots,T$}{
$ D_t\leftarrow \{ u\in V:R(1+\rho_t)\in [d(v_t,u)-x_u,d(v_t,u)+x_u]\} \setminus (\cup _{j=1}^{t-1}B_j) $.\\
$ D\leftarrow D\cup D_t$.\\
$ B_t\leftarrow B(v_t,R(1+\rho _t)) \cap V$.\\
$ S_t \leftarrow B_t\setminus (D\cup ( \cup _{j=1}^{t-1}B_j))$.
}
\Return{$D$ and $S_1,\ldots,S_T$.}
\end{algorithm}

\begin{figure}[!ht]
  \centering
    \includegraphics[width=\linewidth]{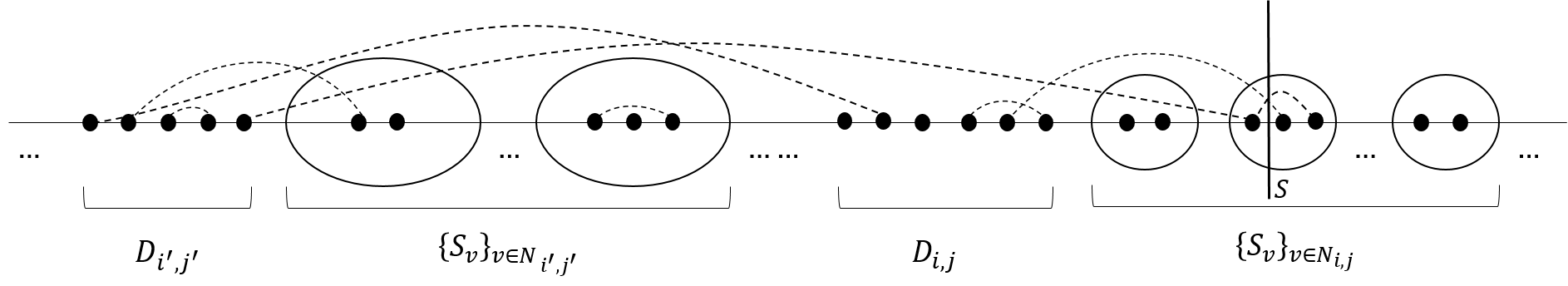}
    \caption{The order of pieces in Algorithm \ref{alg:pathwidth}. The vertical line depicts a cut; each edge crossing it belongs to $ G[S]$ or touches some $ w\in D_{j'}$ for $ j'\leq j$.}
    \label{fig:pathwidth}
\end{figure}

Second, we describe the modifications we make in Algorithm \ref{alg:cutwidth}, which are only in Phase $3$ of the algorithm (see Algorithm \ref{alg:pathwidth}).
The modifications are the following two: $(1)$ Algorithm \ref{alg:cutting-vertex} (via Lemma \ref{lem:cutting-vertex}) is used instead of  Lemma \ref{lem:cutting} (which uses Algorithm \ref{alg:cutting}), and we denote the set $D$ returned in iteration $(j,i)$ of Phase 3 as $D_{i,j}$; and $(2)$ for every vertex $ w\in D_{i,j}$ a singleton piece $ \{ w\}$ is created and all these singleton pieces are arranged before the $ \{S_v\} _{v\in N_{i,j}}$ pieces.
We note that the order in which Algorithm \ref{alg:cutting-vertex} is executed remains the same as in Algorithm \ref{alg:cutwidth}, i.e., an increasing order of the size class $j$.
Figure \ref{fig:pathwidth} depicts how Algorithm \ref{alg:pathwidth} orders the pieces.

\RestyleAlgo{ruled}
\begin{algorithm}[hbt!]
\caption{Modified Phase $3$ for Pathwidth}\label{alg:pathwidth}
\texttt{/* phase $3$ - decomposing and ordering */}\\
Set  $V' \leftarrow V$.\\
 \For{$ j=2,\ldots,\ell+1$}{
 \For{$ i=2,\ldots,\ell+1$}{
 Apply Lemma \ref{lem:cutting-vertex} with parameters $(G[V'],N_{i,j},d,2\Delta _i)$ to obtain $D_{i,j}$ and $\{ S_v\}_{v\in N_{i,j}}$.\\
$V' \leftarrow V' \setminus (D_{i,j}\cup (\cup_{v\in N_{ij}} S_v))$\\
Arrange first a piece $ \{ w\}$ for every $w\in D_{i,j}$ and then all pieces in $ \{ S_v\}_{v\in N_{i,j}}$.
 }
}
\end{algorithm}

\subsection{Analysis}
First, we make some crucial observations about the modified cutting procedure (Algorithm \ref{alg:cutting-vertex}).
\begin{lemma}\label{lem:pathwidthNoEdges}
For every $S_t$ in the output of Algorithm \ref{alg:cutting-vertex} and every edge $ e=(u,v)\in E$ with $u\in S_t$, it must be that $v\in S_t$ or $v\in D$.
\end{lemma}
\begin{proof}
First, assume to the contrary that $ v\in S_{t'}$ for some $t'\neq t$.
Without loss of generality assume that $ t<t'$.
 As $u \in S_{t}$ it must have not been included in $D_t$ when $S_t$ was created, and thus
it must be that $R(1+\rho_t)> d(v_t, u) +x_u$. Similarly, as $v \in S_{t'}$, and $t'>t$, it must not lie in $D_t$ and thus $R(1+\rho_{t})< d(v_{t}, v) - x_v$. 
Hence, this  gives that $x_u +x_v < d(v_t,v) - d(v_t,u) \leq d(u,v)$, contradicting the  constraint that $x_u +x_v \geq d(u,v)$ in $(\mathrm{LP}_{\mathrm{VS}})$ (constraint (\ref{LPVS:edge})).

Second, assume to the contrary that $ v\in V\setminus (D\cup (\cup _{t\in [T]}S_t))$.
Similarly to before, since $v\notin D_t$ and $ v\notin S_t$ it must be that: $ R(1+\rho_t)<d(v_t,v)-x_v$.
This results in the same contradiction.
\end{proof}
An immediate corollary of the above lemma is that for every two pieces $S_t$ and $S_{t'}$ (where $ t\neq t'$) $ E(S_t,S_{t'})=\emptyset$, i.e., there are no edges between $S_t$ and $S_{t'}$.


Next, we bound the size of $D$ 
by an essentially identical argument to Lemmas \ref{lem:cutting} and \ref{lem:auxCutting}.

\begin{lemma}\label{lem:cutting-vertex}
Given an undirected graph $ G=(V,E)$, a set of $T$ terminals $ \{ v_1,\ldots,v_T\}$, a (semi) metric $(V\cup \{ v_1,\ldots,v_T\},d)$, non-negative vertex values $\{ x_u\}_{u\in V} $, and a radius parameter $ R>0$,
there is an efficient algorithm that
returns disjoint sets $D\subseteq V$ and $S_1,\ldots,S_{T}\subseteq V$ satisfying the following properties (with probability $1$):
\begin{enumerate}
\item $ |D| \leq \frac{8\ln{(2T)}}{R}\sum _{u\in V} x_u$. \label{decompose-vertex2}
\item $ S_t\subseteq B( v_t,2R) \cap V$ for every $ t\in [T]$.\label{decompose-vertex1}
\item $D\cup \left( \cup _{t\in [T]}S_t\right)\supseteq \left( \cup _{t\in [T] }B(v_t,R)\right)\cap V $\label{decompose-vertex3}
\end{enumerate}
\end{lemma}
\begin{proof}
The only difference now is that instead of bounding the probability of an edge being cut (Lemma \ref{lem:auxCutting}), we need to bound the probability of a vertex $u$ belonging to $D$.
We will show that this probability is at most $2x_u\cdot \ln (2T)/R$. 


We say that $u$ is settled by terminal $v_t$ if $v_t$ is the first terminal for which: $ d(v_t,u)-x_u\leq R(1+\rho _t)$.
Moreover, we say that $u$ is cut by terminal $ v_t$ if $v_t$ settles $u$ and $ R(1+\rho_t)\leq d(v_t,u)+x_u$.
We note that $u\in D$ if and only if there exists a terminal $v_t$ that cuts $u$.
As in the proof of Lemma \ref{lem:auxCutting},
let $p_t = \Pr[v_t \text{ settles }  u | v_1, \ldots,v_{t-1} \text{ did not settle } u]$ and $q_t= \Pr[u \text{ is cut by }  v_t | v_t \text{ settles } u]$. Then, as in the proof of Lemma \ref{lem:auxCutting} we have: 
    $\Pr[u\in D] \leq  \max_{t \in [T]} q_t$.
Now, since: 
\[q_t = \Pr[R(1+\rho_t) < d(v_t,u)+x_u | R(1+\rho_t) \geq d(v_t,u)-x_u],\] 
we have that $q_t \leq 2x_u \ln (2T)/R$ exactly as in the proof of Lemma \ref{lem:auxCutting}. From this, the claimed bound of requirement (\ref{decompose-vertex2}) follows exactly by repeating the argument in Lemma \ref{lem:cutting}.
Moreover, requirements (\ref{decompose-vertex1}) and (\ref{decompose-vertex3}) also follow by repeating the argument in Lemma \ref{lem:cutting}.
\end{proof}

\paragraph{Bounding the Overall Pathwidth.} 
We now bound the overall VS number (and hence pathwidth) of our algorithm.
Let $ D_j\triangleq \cup _{i=2}^{\ell+1}D_{i,j}$. 
First, we note that as in Lemma \ref{lem:bound-size} the sets $D_j$ for $j=2,\ldots,\ell+1$ and $S_v$ for $v \in N$
give a partition of $V$.

Second, fix a position $r\in [n]$. 
We need to bound the number of vertices at positions $1$ to $r$ that have neighbors at positions $r+1$ to $n$. 
Suppose first that the vertex at position $r$ belongs to some piece $S_v$, where $ v\in N_{i,j}$.
Then by Lemma \ref{lem:pathwidthNoEdges},
the number of such (crossing) vertices is at most  $\sum _{j'=2}^j |D_{j'}|$, plus the VS number of the ordering produced by the algorithm for $ G[S_v]$ recursively (see Figure \ref{fig:pathwidth}).
Otherwise, if some vertex $w \in D_{i,j}$ is placed at position $r$, then we have an even better
upper bound 
of only $\sum _{j'=2}^j |D_{j'}|$.

This gives the following recursive relation that upper bounds $ \mathsf{Alg}(G)$ as follows:
\begin{align} \mathsf{Alg}(G)\leq \max _{j=2,\ldots,\ell+1} \left( \sum _{j'=2}^{j}|D_{j'}| + \max _{v\in N_j}\mathsf{Alg}(G[S_v])\right).\label{pathwidth-recursion-alg}
\end{align}
We will show that for every $ j$ and every $ S_v$, where $v\in N_j$, the bound $ \sum _{j'=2}^j |D_{j'}|\leq 8^{\ell +4}\gamma^2 \cdot \log{(n/|S_v|)}\cdot \mathrm{VS}(G)$.
The analysis then follows directly as in Section \ref{sec:cutwidth}.
We now give the details.
\begin{proof}[Proof of Theorem \ref{thrm:VS}]
We note two things.
First, $ |N_{i,j}|\leq n/n_j=2^{\gamma^{j-1}}$ for every $i$ and $j$ as in the proof of Lemma \ref{lem:pass-over}.
Second, for every $ S_v$, where $v\in N_j$: $|S_v|\leq n_{j-2}$ (as the proof of the second part of Lemma \ref{lem:bound-size} applies here as well).
Together this gives the following bound for every $j$:
\begin{align}
   |D_j|  =\sum _{i=2}^{\ell+1}|D_{i,j}| & \leq 8\cdot \frac{P^*}{n}\sum _{i=2}^{\ell+1} 8^i\cdot \ln{(2|N_{i,j}|)} \tag{by Lemma \ref{lem:cutting-vertex}, $ P^*=\sum _{u\in V}x_u$ and $ \Delta _i=n/8^i$}\\
   & \leq 32\cdot \frac{P^*}{n}\cdot \gamma ^{j-1}\sum _{i=2}^{\ell +1}8^i \tag{as $|N_{i,j}|\leq n/n_j$ and $ \log{(n/n_j)}=\gamma ^{j-1}$}\\
   & \leq 8^{\ell +3}\cdot\mathrm{VS}(G)\cdot \gamma ^{j-1}. \tag{as $P^* \leq n \cdot\mathrm{VS}(G) $}
\end{align}
This gives that 
\[
\sum _{j'=2}^j|D_{j'}|
 \leq 8^{\ell+4}\cdot \mathrm{VS}(G)\cdot \gamma^{j-1}
 \leq 8^{\ell+4}\gamma^2\cdot \mathrm{VS}(G) \cdot \log{(n/|S_v|)},\]
where the first inequality uses that $ \gamma+\gamma^2+\ldots + \gamma ^{j-1}\leq 2\cdot \gamma^{j-1}$
and the second equality uses that $ |S_v|\leq n_{j-2}$
for every $ S_v$ with $v\in N_j$ (and that $ \gamma ^{j-3}\leq \log{(n/n_{j-2})}$).
The rest of the proof proceeds as in Section \ref{sec:cutwidth} and this completes the proof of Theorem \ref{thrm:VS}.
\end{proof}
\section{Open Problems and New Directions}
\label{sec:discussion}

We now discuss some new directions and open problems suggested by our work.

A natural question is to further improve our  approximation ratios for cutwidth and pathwidth.
In Section \ref{sec:disc-local}, we
propose two natural problems, that can be viewed as extensions of Seymour's cutting lemma, and
show that if an affirmative answer to either one would imply an $O(\log n\log \log n)$-approximation for both cutwidth and pathwidth. We believe the proposed questions are also interesting on their own.

In Section \ref{sec:LPrelaxations}, we give direct LP relaxations for cutwidth and pathwidth. These are based on the idea of flow metrics due to Bornstein and Vempala \cite{BV04}, and to the best of our knowledge no direct relaxations were known previously for both cutwidth and pathwidth (for simplicity of presentation in what follows we focus on cutwidth).
Remarkably, this relaxation directly gives a metric for every induced subgraph $G[S]$ of $G$. 
Recall that in contrast, in our current approach the LP for MLA only gives a metric for $G$, and we partition $V$ into the pieces $S_1,\ldots,S_k$ without any knowledge of the metric on the induced subgraphs $G[S_i]$.
Moreover, we show that this new direct LP relaxation provides a lower bound on cutwidth that is at least as large as $ \Omega (\max _{S\subseteq V:S\neq \emptyset}\mathrm{L}^*(G[S])/|S|)$.
Thus, it is plausible that this new and stronger LP can lead to better approximation guarantees.

In Section \ref{sec:directSDP}, 
we give SDP relaxations for cutwidth and pathwidth based on $\ell_2^2$ metrics. 
This is achieved by extending the flow metric approach of \cite{BV04} from LP to SDP (which may be of independent interest).
Potentially, these SDP relaxations may allow $\log^{1/2 + o(1)}(n)$ approximations for cutwidth and pathwidth.

Finally, in Section \ref{sec:DirectedLAyout} we address $\min$-$\max$ graph layout problems on directed graphs.
We briefly describe why the natural adaptations of our approach fail for directed graphs, and propose two natural min-max layout problems in the directed setting that are open.

\subsection{Improving the $\beta(n)=\log^{o(1)}(n)$ Term via Local Partitions}
\label{sec:disc-local}
We propose an approach to improve the $ \beta(n)=\log^{o(1)}(n)$ term in the approximation to $O(\log \log n)$.
This approach extends Seymour's cutting lemma (see \cite{seymour1995packing,even2000divide}) in a natural way, and 
we call it {\em local partition}.
This question is interesting on its own, and we believe should have additional applications beyond those studied in this paper.

Seymour's cutting lemma, when applied to cutwidth, says the following. 
Given a graph $G=(V,E)$ on $n$ vertices and a spreading metric $d$ on $V$, one can efficiently 
find a subset $S\subseteq V$ satisfying $|S|\leq n/2$ and,
\[ \left|\delta(S) \right|\leq \frac{\text{vol}(S)}{n}\cdot \log\left( e\cdot\frac{\text{vol}(V)}{\text{vol}(S)}\right)  O\left(\log\log n\right).\]
Here, the volume of an edge $ (u,v)\in E$ is the distance $d(u,v)$,
and the volume of a subset $S\subseteq V$ is $ \text{vol}(S)\triangleq \sum _{(u,v)\in E(S)}d(u,v)$.\footnote{The volume $\text{vol}(S)$ includes partial volumes of edges crossing the boundary of $S$ plus an additional seed value. However, we ignore these technicalities here for simplicity of presentation.}
By standard edge reweighting one also has $ \log\left(\text{vol}(V)\right)=O(\log {n})$.

\vspace{2mm}

Notice that the bound above does not give any control on the other piece $V\setminus S$.
The following notion, that we call {\em local partition}, addresses this issue. 

\vspace{2mm}

\begin{definition}[Local Partition] A partition $\{S_1,\ldots,S_k\}$ of $V$ is a local partition if each $i \in [k]$, we have $|S_i|\leq n/2$ and, 
\begin{equation}
    \label{eq:local-partition}
\left|\delta(S_i)\right| \leq  \frac{\mathrm{vol}(S_i)}{n} \cdot \log\left( e\cdot\frac{\mathrm{vol}(V)}{\mathrm{vol}(S_i)}\right)  O\left(     \log \log n \right).
\end{equation}
\end{definition}

In Theorem \ref{thm:partition-to-cw} below we show that a procedure to find a local partition will give a $O(\log n\log \log n)$ approximation for cutwidth.
In fact, it shows that the following weaker notion also suffices for cutwidth.

 \begin{definition}[Weak Local Partition] 
A partition $\{S_1,\ldots,S_k\}$ of $V$ is a weak local partition, if for each $i\in [k]$ we have $|S_i|\leq n/2$ and, 
\begin{equation}
    \label{eq:weak-local-partition}
\sum_{j: \mathrm{vol}(S_j) \geq \mathrm{vol}(S_i)} |\delta(S_j)| \leq \frac{\mathrm{vol}(V)}{n} \cdot \log \left( e\cdot\frac{\mathrm{vol}(V)}{\mathrm{vol}(S_i)}\right)  O\left( \log \log n \right).    
\end{equation}
\end{definition}
Notice that 
a local partition is also a weak local partition. Indeed, to prove \eqref{eq:weak-local-partition} for any $ i\in [k]$, sum \eqref{eq:local-partition} over all $j$ such that $\text{vol}(S_j) \geq \text{vol}(S_i)$ and use that (i) $\log (\text{vol}(V)/\text{vol}(S_j)) \leq \log (\text{vol}(V)/\text{vol}(S_i))$  
and (ii) $\sum _{j:\text{vol}(S_j)\geq \text{vol}(S_i)}\text{vol}(S_j)\leq \sum _{j=1}^k \text{vol}(S_j)\leq \text{vol}(V)$.

\begin{theorem}
\label{thm:partition-to-cw}
    An efficient procedure to find a weak local partition implies an efficient $O(\log n \log \log n)$ approximation for \CW.
\end{theorem}
\begin{proof}(Sketch)
Consider the following algorithm.
Given the graph $G$, solve the LP for MLA in Section \ref{sec:preliminaries} to obtain the metric $d$. 
Consider the weak local partition $\{S_1,\ldots, S_k\}$ obtained from $(V,d)$ as guaranteed in \eqref{eq:weak-local-partition}.
Order the pieces $\{ S_1,\ldots,S_k\}$ in a non-increasing order of volume, i.e., $ \text{vol}(S_1)\geq \text{vol}(S_2)\geq \ldots \geq \text{vol}(S_k)$, and recurse on each piece $S_i$.


Let $C$ denote the optimum cutwidth and $\mathrm{Alg}(G)$ denote the cutwidth of the algorithm on $G$. Then,
\begin{equation}
    \mathrm{Alg}(G) \leq \max_{i \in [k]}  \Big( \mathrm{Alg}(G[S_i])  + \sum_{j:j\leq i} |\delta(S_{j})| \Big).\label{overview-recurrence}
\end{equation}
This holds for the same reason as in Section \ref{sec:cutwidth-analysis}. Consider some vertex $ v\in S_i$ in the final ordering. The number of edges crossing $v$ is at most
(i) the cutwidth of $ G[S_i]$ produced by the algorithm when it recurses on $ S_i$, plus (ii) the number of edges between two different pieces $S_{\ell}$ and $S_r$ that cross $v$, which is at most $\sum _{j:j\leq i}|\delta (S_j)|$.

As $ \text{vol}(V)/n = O(C)$, and $\{S_1,\ldots,S_k\}$ is a weak-local partition, \eqref{eq:weak-local-partition} gives that for each $i \in [k]$,
\begin{align}
\sum _{j:\text{vol}(S_j)\geq 
\text{vol}(S_i)}\left| \delta (S_j)\right| 
 \leq C \cdot \log \left( e \cdot\frac{\text{vol}(V)}{\text{vol}(S_i)}\right) O(\log{\log{n}}).\label{eq:LocalPartition2}
\end{align}
Plugging \eqref{eq:LocalPartition2} in \eqref{overview-recurrence}, and (i) assuming inductively that $
\mathrm{Alg}(G[S]) \leq c\cdot C\cdot \log{ (\text{vol}(S))} O(\log{\log{n}})$ (for some large enough absolute constant $c$) for every $S$ with $|S|<n$; and (ii) recalling that $ |S_i|\leq n/2$, we get that $\mathrm{Alg}(G) \leq c\cdot C\cdot \log{ (\text{vol}(V))} O(\log{\log{n}})$.
\end{proof}
Thus, we propose the following open questions.
\begin{question}
Does a local partition always exist and can it be computed efficiently?
\end{question}

\begin{question}
Does a weak local partition always exist and can it be computed efficiently?
\end{question}

Analogously, one can define the node counterpart of local partitions  and weak local partitions, and show that they imply an $ O(\log{n}\log{\log{n}})$ approximation for pathwidth.

\subsection{Direct LP Relaxations using Flow Metrics}  \label{sec:LPrelaxations}

We focus on cutwidth, as the arguments for pathwidth are similar and thus deferred to a full version of the paper. 
Given a graph $G$, recall that $\mathrm{CW}(G)$ denotes the optimal cutwidth. Let us denote $C^* = \max_{S\subseteq V:S\neq \emptyset} \mathrm{L}^*(G[S])/|S|$, while recalling that $\mathrm{L}^*(G)$ is the optimum value of the LP relaxation for MLA on $G$ (as defined in Section \ref{sec:preliminaries}). Recall that even though we can compute $\mathrm{L}^*(G[S])$ for any given $S$, it is not clear how to compute the quantity $C^*$,
as it involves taking the maximum over all possible $S\subseteq V$.

We now give a new direct LP relaxation for cutwidth, which we denote by $ \mathrm{LP_{CW}}$, and show the following key result.
    \begin{theorem}
    \label{thm:strong-lb}
    The optimum value $C^{**}$ of $\mathrm{LP_{CW}}$ (described below) is $\Omega(C^*)$.
\end{theorem}
\paragraph{Direct LP Relaxation for Cutwidth.} 
The idea is to define a (semi)-metric $d_x$ on $V$ for every vertex $x\in V$, using the variables $ d_x(y,z)$ for every pair of vertices $ y,z\in V$.
In the intended integral solution, $d_x$ is the cut metric separating vertices to the  left of $x$ (including $x$) from those to the right of $x$.
I.e., $ d_x(y,z)=1$ if $ |S_x\cap \{ y,z\}|=1$ and $ d_x(y,z)=0$ otherwise (where $S_x$ consists of all vertices to the left of $x$ including $x$).

Consider the following LP.
\begin{align}
(\mathrm{LP_{CW}})~~~~~~~\min~~~ &  C \notag \\ 
\text{s.t.}~~~& d_x(y,z)+d_x(z,w) \geq d_x(y,w) & \forall y,z,w\in V, \forall x\in V \label{LPnew:triangle} \\
& d_x(y,z)=d_x(z,y) & \forall y,z\in V, \forall x\in V \label{LPnew:symmetry}\\
& d_x(y,y)=0 & \forall y\in V, \forall x\in V \label{LPmew:zerodist}\\
& d_x(y,z)+d_y(x,z)+d_z(x,y)\geq 1 & \forall \{ x,y,z\}\subseteq V \label{LPnew:spreading3}\\
& d_x(x,y)+d_y(x,y)\geq 1 & \forall \{ x,y\}\subseteq V \label{LPnew:spreading2}\\
& \sum _{(u,v)\in E}d_x(u,v) \leq C & \forall x\in V \label{LPnew:cutwidth}\\
& d_x(y,z)\geq 0  & \forall y,z\in V , \forall x\in V \nonumber 
\end{align}
To see why this is a valid relaxation for cutwidth, consider some (integral) linear ordering $\pi$ of $G$. Constraints (\ref{LPnew:triangle}),(\ref{LPnew:symmetry}), and (\ref{LPmew:zerodist}) impose
that $d_x$ is a metric.
Constraint (\ref{LPnew:spreading3}) holds as for any three vertices $ x,y$ and $z$, one of them must lie between the other two in any ordering.
Similarly, (\ref{LPnew:spreading2}) holds as for any pair of vertices $ \{ x,y\}\subseteq V$ one of them must lie to the left of the other.
Constraint (\ref{LPnew:cutwidth}) holds as $C$ is at least as large as the number of edges  crossing $x$ in the ordering $ \pi$, which is the volume of each $d_x$ metric.

\paragraph{Induced Metric on $G[S]$.}
Consider some optimal solution $\{d_x(y,z)\}_{x,y,z\in V}$ to $\mathrm{LP_{CW}}$ and let $C^{**}$ denote the objective value of this solution. For any subset $S\subseteq V$ 
let us define:
\[d_S(y,z)\triangleq \sum _{x\in S} d_x(y,z) \qquad  \text{for  each $ y,z\in V$}.\]
As each $d_x$ is a metric, it directly follows that $d_S$ is also a metric on $S$.
More importantly, we have the following crucial property.
\begin{lemma}
\label{lem:ds-feasible}
There is an absolute constant $c>0$, such that 
   for every subset $S \subseteq V$, the metric $c\cdot d_S$ is  a feasible solution to the LP for MLA on $G[S]$.  
\end{lemma}
\noindent {\em Remark:} Lemma \ref{lem:ds-feasible} implies that our algorithm for cutwidth (Section \ref{sec:cutwidth}) can use $\mathrm{LP_{CW}}$ as follows: it solves $ \mathrm{LP_{CW}}$ once for $G$ and when it recurses on $ G[S]$ it uses $ c\cdot d_S$ (instead of using the metric obtained from solving the LP for MLA on $ G[S]$).

Let us first see why Lemma \ref{lem:ds-feasible} immediately implies Theorem \ref{thm:strong-lb}.
\begin{proof}[Proof of Theorem \ref{thm:strong-lb}.]
Fix some non-empty subset of vertices $S$. By Lemma \ref{lem:ds-feasible}, as the metric $c \cdot d_S$ is a feasible solution to the LP relaxation for MLA on $G[S]$, we have:  
\[  \mathrm{L}^*(G[S]) \leq   \sum _{(y,z)\in E(S)} c \cdot d_S(y,z)  \leq \sum_{(y,z)\in E} c \cdot d_S(y,z) \leq c |S| C^{**},\] 
where the last inequality follows as $C^{**}\geq \sum_{(y,z)\in E}d_x(y,z) $ for each $x$ by constraint \eqref{LPnew:cutwidth}.
Thus, $C^{**} \geq \mathrm{L}^*(G[S])/(c|S|)$ for every $S\subseteq V$ and hence $C^{**} = \Omega(C^*)$.
 \end{proof}

To prove Lemma \ref{lem:ds-feasible}, we first prove the following lemma.
\begin{lemma}\label{lem:newLP-spread}
$\sum _{y,z\in W} d_W(y,z)\geq \Omega(|W|^3)$ for every $ W\subseteq V$ of size $ |W|\geq 2$.
\end{lemma}
\begin{proof}
If $ |W|=2$ the claim follows from constraint (\ref{LPnew:spreading2}). If $ |W|\geq 3$ then,
\begin{align*}
\sum _{y,z\in W}d_W(y,z) = \sum _{y,z\in W}\sum _{x\in W}d_x(y,z) \geq \sum _{\{ x,y,z\}\subseteq W}\left( d_x(y,z)+d_y(x,z)+d_z(x,y)\right) = \Omega(|W|^3),
\end{align*}
where the last step follows from constraint (\ref{LPnew:spreading3}).
\end{proof}

\begin{proof}[Proof of Lemma \ref{lem:ds-feasible}.]
Fix some subset $S\subseteq V$ of vertices, and consider the LP relaxation for MLA on $G[S]$.
It suffices to show that there exists an absolute constant $c'>0$ such that $d_S$ satisfies the spreading constraints 
$ \sum _{z\in T}d_S(y,z)\geq \frac{c'}{4}(|T|^2-1)$
for every vertex $y \in S$ and every subset $T \subseteq S$. 

We assume that $|T|\geq 2$, otherwise the statement is trivially true.
Suppose for the sake of contradiction that $ \sum_{z \in T} d_S(y,z) < c' (|T|^2-1)$.
Consider the set of vertices $ W= \{ z\in T: d_S(y,z)\leq c'(|T|+1)\}$.  
By Markov's inequality, $|W| \geq 3|T|/4+1/4$ (and as $|W|$ is integral, this also means that $|W|\geq 2$). Moreover, by triangle inequality $d_S(z,z') \leq 2 c'(|T|+1)$ for all $z,z'\in W$.
But for $c'$ small enough absolute constant, this contradicts Lemma \ref{lem:newLP-spread} since:   \[ \sum_{z,z'\in W} d_W(z,z')  \leq  \sum_{z,z'\in W} d_S(z,z')
 \leq |W|^2 \cdot 2c'(|T|+1) \leq 4 c'|W|^3. \qedhere\]
\end{proof}


Given Theorem \ref{thm:strong-lb}, it may be plausible that $\mathrm{LP_{CW}}$ gives a potentially much stronger lower bound on cutwidth, than the one we use in our algorithm in Section \ref{sec:cutwidth}.

This leads to the following  two interesting open questions.

\begin{question}
Can $\mathrm{LP_{CW}}$ be exploited to achieve an approximation better than $ O(\beta(n)\cdot\log{n})$ for cutwidth? Perhaps even $O(\log n)$ or $O(\log n \log \log n)$? In particular, can we use $\mathrm{LP_{CW}}$'s solution and the information from the metrics $d_S$ for every possible $S$, to construct a better partition of $V$ 
as discussed in Section \ref{sec:disc-local}?
\end{question}

Second, even though we do not know how to compute $C^*$ efficiently, it would be interesting to understand whether $C^{**}$ can be substantially better than $C^*$. This may lead to useful structural insights on how the more global information given by $\mathrm{LP_{CW}}$ can be exploited.
\begin{question} Are there graphs for which  $C^{**} =\omega(C^*)$? 
 \end{question}

Finally, we remark that one can write a similar direct LP relaxation for pathwidth, by considering node cuts instead of edge cuts. 
Here, we can define $d_S$ as above and analogous versions of Theorem \ref{thm:strong-lb} and Lemma \ref{lem:newLP-spread} hold as well.

\subsection{Direct SDP Relaxations using Flow Metrics and $\ell_2^2$ Metrics}\label{sec:directSDP}
We observe that one can formulate a direct SDP relaxation for cutwidth and pathwidth.
This might allow one to obtain a better approximation by utilizing the machinery of Arora, Rao, and Vazirani \cite{arora2009expander}.
For example, this was done for MLA by \cite{feige2007improved,charikar2010ℓ} who presented an approximation of $O(\log^{1/2}{n}\log{\log{n}})$ inspired by a framework of \cite{rao2005new} for approximating MLA in planar graphs.

We consider cutwidth below.
The relaxation defines an $\ell_2^2$ metric over $V$ for every vertex $x\in V$ by the unit vectors $ \{\bu_y^x\}_{y\in V}$.
This is captured by constraints (\ref{SDP:unit}) and (\ref{SDP:triangle}).
In the intended integral solution the $\ell_2^2$ metric induced by $ \{ \bu^x_y\}_{y\in V}$, i.e., $ \frac{1}{4}\| \bu^x_y - \bu^x_z\|_2^2$, is the distance between $y$ and $z$ in the cut metric for $x$ separating vertices placed to its left (including $x$) from those to its right.
Constraint (\ref{SDP:orthogonal}) ensures that the vectors representing these $\ell_2^2$ metrics are orthogonal over the different metrics, i.e., if $ x\neq z$ then $ \bu _y^x \cdot \bu _w^z = 0$ for any two (not necessarily different) vertices $ y,w\in V$.
The spreading constraints are captured by (\ref{SDP:spreading3}) and (\ref{SDP:spreading2}), and the constraint (\ref{SDP:cutwidth}) captures the cutwidth.
\begin{align}
(\mathrm{SDP_{CW}})~~~~~~~\min~~~ &  C \notag \\ 
\text{s.t.}~~~& \bu_y^x \cdot \bu_y^x =1 & \forall y\in V, \forall x\in V \label{SDP:unit}\\
& \|\bu_y^x - \bu_z^x \|_2^2 + \|\bu_z^x - \bu_w^x \|_2^2 \geq \|\bu _y^x - \bu_w^x\|_2^2 & \forall z,y,w\in V, \forall x\in V \label{SDP:triangle}\\
& \bu_y^x \cdot \bu_w^z = 0& \forall y,w\in V, \forall \{ x,z\}\subseteq V\label{SDP:orthogonal}\\
& \frac{1}{4}\|\bu_y^x-\bu_z^x \|_2^2 + \frac{1}{4}\|\bu_x^y - \bu_z^y \|_2^2 + \frac{1}{4}\| \bu_x^z - \bu_y^z\|_2^2 \geq 1& \forall \{ x,y,z\}\subseteq V\label{SDP:spreading3}\\
& \frac{1}{4}\| \bu_x^x - \bu_y^x\|_2^2 + \frac{1}{4}\|\bu_x^y - \bu_y^y \|_2^2 \geq 1 & \forall \{ x,y\}\subseteq V\label{SDP:spreading2}\\
& \sum _{(y,z)\in E}\frac{1}{4}\| \bu_y^x - \bu_z^x\|_2^2 \leq C& \forall x\in V\label{SDP:cutwidth}
\end{align}

Clearly $ (\mathrm{SDP_{CW}})$ is a relaxation for cutwidth.
Let $ \pi$ be some (integral) linear ordering of $G$, and let $\{ \be^x\} _{x\in V}$ be orthonormal vectors one for each $ x\in V$.
As before, consider the cut $ S_x\subseteq V$ consisting vertices to the left of $x$ in $ \pi$ including $x$ itself.
For every $ x\in V$ define $ \bu_y^x=\be^x$ if $ y\in S_x$ and $ \bu_y^x=-\be^x$ if $ y\notin S_x$.
So the $\ell_2^2$ metric induced by $ \{ \bu^x_y\} _{y\in V}$, i.e., $ \frac{1}{4}\| \bu^x_y - \bu^x_z\|_2^2$ is the distance between $y$ and $z$, in the cut metric of $ S_x$.
Clearly, constraints (\ref{SDP:unit}), (\ref{SDP:triangle}), and (\ref{SDP:orthogonal}) hold.
Constraint (\ref{SDP:spreading3}) holds as for any three vertices $x,y$ and $z$ one of them must be placed between the other two.
Similarly,  (\ref{SDP:spreading2}) holds as for any two vertices $x,y$ one must lie to the left of the other.
Finally, (\ref{SDP:cutwidth}) ensures that $C$ is at least the largest number of edges that cross any point.

As before, $ (\mathrm{SDP_{CW}})$ naturally defines an $\ell_2^2$ metric for every subgraph $ G[S]$ induced by $S\subseteq V$ as follows.
Define the vectors
\begin{align}
 \bu^S_y\triangleq \sum _{x\in S}\bu_y^x   \label{SDP:vectors}
\end{align}
for every $ y\in V$.
We claim that $ \{ \bu^S_y\}_{y\in S}$ induces an $ \ell_2^2$ metric over $S$.

\begin{lemma}\label{lem:SDPmetric}
    For every $ S\subseteq V$ the vectors $ \{ \bu_y^S\}_{y\in V}$ satisfy that for every $ y,z,w\in V$,
\[ \|\bu_y^S - \bu_z^S \|_2^2 + \|\bu_z^S - \bu_w^S \|_2^2 \geq \| \bu _y^S - \bu_w^S\|_2^2.\]
\end{lemma}
\begin{proof}
By the orthogonality of $\bu_y^x$ and $\bu^z_w$ whenever $x\neq z$, every vector in $\text{span}_{y\in V}\{\bu_y^x\}$ is orthogonal to every vector in $\text{span}_{w\in V}\{\bu_w^z\}$. Thus we have,
\begin{align*}
     \| \bu_y^S - \bu_z^S\|_2^2 +\|\bu_z^S - \bu_w^S \| _2^2    & = 
       \| \sum_{x\in S} (\bu_y^x-\bu_z^x) \|_2^2 + \|\sum_{x\in S} (\bu_z^x-\bu_w^x) \|_2^2   \nonumber\\
       & = \sum_{x \in S}  \left(\|  \bu_y^x-\bu_z^x \|_2^2 + \|\bu_z^x-\bu_w^x \|_2^2 \right) 
     \geq  \sum _{x\in S} \| \bu_y^x-\bu_w^x\|_2^2 \tag{$\ell_2^2$ triangle inequality} \\
     & = \| \sum _{x\in S}  (\bu_y^x-\bu_w^x)\|_2^2  = \| \bu_y^S - \bu_w^S\|_2^2. \qedhere
\end{align*}
\end{proof}

As before, we have that for every subgraph $G[S]$,
\begin{align}
\frac{1}{|S|}\sum _{(y,z)\in E(S)}\frac{1}{4}\|\bu_y^S - \bu_z^S \|_2^2 \leq \frac{1}{|S|}\sum _{(y,z)\in E}\frac{1}{4}\| \bu^S_y-\bu^S_z\|_2^2 \leq C,  \label{SDP:lowerbound}
\end{align}
and moreover,
constraints (\ref{SDP:spreading3}) and (\ref{SDP:spreading2}) imply the following spreading property (Lemma \ref{lem:SDPradius}), which implies a lemma similar to Lemma \ref{lem:ds-feasible}. 

\begin{lemma}\label{lem:SDPradius}
    $ \sum _{y,z\in S}\| \bu_y^S - \bu_z^S\|_2^2 \geq \Omega(|S|^3)$ for every $S\subseteq V$ with $ |S|\geq 2$.
\end{lemma}
\begin{proof}
If $ |S|=2$ the claim follows from constraint (\ref{SDP:spreading2}). If $ |S|\geq 3$ then,
\begin{align*}
\sum _{y,z\in S}\| \bu_y^S - \bu_z^S\|_2^2 = & \sum _{y,z\in S}\sum _{x\in S}\| \bu_y^x - \bu_z^x\|_2^2\label{eq:SDPspreading1} 
\geq  \sum _{x,y,z\subseteq S}\left( \| \bu_y^x-\bu_z^x\|_2^2+\| \bu_x^y-\bu_z^y\|_2^2+\| \bu_x^z-\bu_y^z\|_2^2\right)  = \Omega(|S|^3),
\end{align*}
where the last step 
follows from constraint (\ref{SDP:spreading3}). 
\end{proof}
Thus, we propose the following open question.
\begin{question}
Is there an efficient rounding algorithm for $ (\mathrm{SDP_{CW}})$ that achieves an approximation better than $ O(\beta(n)\cdot \log{n})$?  Perhaps even  $O(\log^{1/2+\epsilon} n)$ for some small $ \epsilon >0$?
\end{question}
Similarly one can also write a direct SDP relaxation for pathwidth, where node cuts are considered instead of edge cuts.

\subsection{Directed Min-Max Layout Problems}\label{sec:DirectedLAyout}
Several interesting graph layout problems involve directed graphs.
Here, typically the input consists of a directed acyclic graph (DAG) and the ordering $\pi$ is required to be a topological ordering.
Many such directed graph layout problems have been studied, e.g., \RS \cite{LR99,ravi1991ordering,WAT14}, \US \cite{even2000divide,LR99,ravi1991ordering}, \FAS \cite{even2000divide,LR99,seymour1995packing}, and \STP \cite{charikar2010ℓ,rao2005new}.
Similarly to undirected graph layout problems, a lot of progress has been made for directed layout problems with $\min$-sum objectives, but there has been essentially no progress for
$\min$-$\max$ objectives.

Perhaps the most well-known directed graph layout problem with a $\min$-$\max$  objective is the \RS problem (RS).
Here given a DAG $ G=(V,E)$, the goal is to find a topological ordering $\pi$ that minimizes the worst node cut.
Formally, let $\Gamma(S^{\pi}_i)$ denote the collection of nodes in $ S^{\pi}_i$ that have a neighbor outside $S^{\pi}_i$ (recall that $ S^{\pi}_i$ is the set of nodes at positions $1$ to $i$ in $\pi$).
The goal in the RS problem is to compute \[ \min _{\pi}\max_{i\in [n]} |\Gamma(S^{\pi}_i)|,\]
where the minimum is taken over all topological orderings $\pi$ of $G$.
Intuitively, RS is the directed counterpart of pathwidth, and the best known approximation for it is $ O(\log^2 n)$ due to Klein, Agrawal, Ravi, and Rao \cite{klein1990approximation}.

Another directed $\min$-$\max$ problem, that intuitively is the directed counterpart of cutwidth, is the \ST (MS) problem \cite{kayaaslan2018scheduling,liu1987application,WAT14}.
The input here is a DAG $ G=(V,E)$, and the goal is to find a topological ordering $ \pi$ that minimizes the worst edge cut, i.e., \[ \min _{\pi} \max _{i\in [n]}|\delta(S^{\pi}_i)|.\]

Interestingly, the approach in this paper does not seem to work, at least in a straightforward way, for such directed layout problems.
The problem is that as the graph is directed, the resulting metric is assymmetric and some of the steps break down. 
Let us discuss this briefly using MS as an example.

Here, one can formulate the directed counterpart of the MLA relaxation, which is the known relaxation for \STP, see, e.g., \cite{rao2005new}.
This provides a suitable directed metric.
In order to adapt our approach, one needs to decide which type of balls to use around each vertex $x\in V$: balls defined by distances {\em from} $x$, i.e., $ B(x,r)=\{ y\in V:d(x,y)\leq r\}$, or distances {\em to} $x$, i.e., $ B(x,r)=\{ y\in V:d(y,x)\leq r\}$.
This is not a problem if one grows a single ball (as in previous works), but since we need to create multiple balls at various difference scales, the net argument and the packing-covering duality breaks down due to assymmetry.

This leads to the following interesting open question.
\begin{question}
Can our approach be extended to directed $\min$-$\max$ graph layout problems? 
In particular, can we find almost logarithmic approximations for \ST or \RS?
\end{question}
\noindent {\em Remark.} As far as we know, nothing better than an $O(\log^2 n)$  approximation is known for these two problems. However, it seems plausible that one can obtain an $O(\log^{1.5} n)$ approximation for them using \cite{arora2009expander} and the techniques of \cite{feige2005improved, AgarwalCMM05}. This would be an interesting first step.

 \section*{Acknowledgment}
 We would like to thank Jens Vygen for suggesting to us the problem of simultaneously approximating the cutwidth and linear arrangement objectives, which led to the results in Section \ref{sec:sim-linear}.
\bibliographystyle{alpha}
{ \bibliography{main}}

\appendix

\section{The Cutting Lemma and Proof of Lemma \ref{lem:cutting}}
\label{apx:CuttingLemmaProof}
Consider Algorithm \ref{alg:cutting} below.  
\RestyleAlgo{ruled}
\begin{algorithm}[hbt!]
\caption{Metric Decomposition $ (G=(V,E),\{ v_1,\ldots,v_{T}\},d,R)$}\label{alg:cutting}
$\lambda \leftarrow \ln{(2T)}$.\\
Sample $ \rho_1,\ldots,\rho_{T}$ i.i.d distributed  $ \text{exp}(\lambda)$ repeatedly, until $ \rho _t\leq 1$ $ \forall t\in [T]$.\\
\For{$ t=1,\ldots,T$}{
$ B_t\leftarrow B(v_t,R(1+\rho _t)) \cap V$.\\
$ S_t \leftarrow B_t\setminus ( \cup _{j=1}^{t-1}B_j)$.
}
\Return{$S_1,\ldots,S_T$.}
\end{algorithm}

We say that an edge $e=(u,v)\in E$ is cut if there exists $ t\in [T]$ such that $e \in \delta(S_t)$. 
In order to prove Lemma \ref{lem:cutting} we need the following auxiliary lemma.
\begin{lemma}\label{lem:auxCutting}
Let   $\mathcal{E}$ be the event $\mathcal{E} = \{\rho_1,\ldots,\rho_T\leq 1\}$. Then $\Pr [\mathcal{E}]\geq \nicefrac{1}{2} $, and for each $e=(u,v)\in E$,
\[ \Pr \left[ e\text{ is cut}\,| \mathcal{E}\right] \leq  (2d(u,v)\cdot \ln(2T))/R. \label{lem:cutting-bd-A}
\]
\end{lemma}

\begin{proof}
Fix an edge $ e=(u,v)\in E$.
We note that the event that $e$ is cut is equivalent to the event that $e$ crosses the boundary of at least one of the $S_t$ sets, i.e., 
\[ \Pr [e\text{ is cut}|\rho_1,\ldots,\rho_T\leq 1]=\Pr [\exists t\in [T] \text{ s.t. } e\in \delta (S_t)|\rho_1,\ldots,\rho_T\leq 1].\]
First, we upper bound the probability of the non-conditioned event that $e$ is separated, i.e., $ \Pr [\exists t\in [T] \text{ s.t. } e\in \delta (S_t)]$.
We say that $ v_t$ settles $e$ if it is the first terminal that chooses at least one of $u$ and $v$, i.e., $ R(1+\rho_t)\geq \min \{ d(v_t,u),d(v_t,v)\}$ and for every $ t'<t$ it holds that $ R(1+\rho_{t'})< \min \{ d(v_{t'},u),d(v_{t'},v)\}$.
Moreover, we say that $ v_t$ cuts $e$ if $v_t$ settles $e$ and $ e\in \delta (S_t)$ (namely that $v_t$ chooses exactly one of $u$ and $v$).
Note that if $ v_t$ settles $e$, then either $ e\in \delta (S_t)$ or $ u,v\in S_t$ (and thus $ e\notin \delta (S_{t'})$ for every $ t'\neq t$.).
Let us denote $ q_t\triangleq \Pr [v_t\text{ cuts }e|v_t\text{ settles }e]$ and $ p_t\triangleq \Pr [v_t\text{ settles }e|v_1,\ldots,v_{t-1}\text{ do not settle }e]$.
Note that:
\begin{align}
  & \Pr [\exists t\in [T] \text{ s.t. } e\in \delta (S_t)]  = \sum _{t=1}^T \Pr [v_t \text{ cuts }e] \label{cutprob1}\\
  & = \sum _{t=1}^T \left(q_t \cdot p_t\cdot \prod _{t'=1}^{t-1}(1-p_{t'})\right)  \leq \max _{t\in [T]} \left( q_t\right) \cdot\sum _{t=1}^T \left( p_t\cdot \prod_{t'=1}^{t-1}(1-p_{t'})\right)\label{cutprob3}   \\ 
  & \leq \max _{t\in [T]} \left( q_t\right) \label{cutprob4}.
\end{align}
Equality (\ref{cutprob1}) follows from the fact that the events $\{ v_t\text{ cuts }e\} $ are disjoint for every $ t\in [T]$. 
Equality (\ref{cutprob3}) follows from the definitions of $ p_t$ and $q_t$.
Finally, inequality (\ref{cutprob4}) follows as $ \sum _{t=1}^T  p_t\cdot \prod_{t'=1}^{t-1}(1-p_{t'}) \leq 1$ (can be proved by induction on $T$).

We now upper bound $ q_t$ as follows (in what follows assume without loss of generality that $ d(v_t,u)\leq d(v_t,v)$):
\begin{align}
q_t &= \Pr [d(v_t,u)\leq R(1+\rho_t)< d(v_t,v) \, | \, R(1+\rho_t) \geq d(v_t,u)] \label{qBound1}\\
& = 1-\Pr [\rho_t\geq (d(v_t,v)-R)/R \, | \, \rho_t\geq (d(v_t,u)-R)/R]\nonumber\\
& = 1-\Pr[\rho_t\geq (d(v_t,v)-d(v_t,u))/R]\label{qBound2}\\
& = 1-\text{exp}(-\lambda\cdot  (d(v_t,v)-d(v_t,u))/R)  \leq \lambda \cdot d(u,v)/R.\label{qBound4}
\end{align}
Equality (\ref{qBound1}) follows from the definition of $q_t$, and (\ref{qBound2}) follows from the memoryless property of the exponential distribution.
Equality (\ref{qBound4}) follows as  $ \rho_t\sim \text{exp}(\lambda)$,
and as $ 1-\text{exp}(-x)\leq x$ for every $x$ and the triangle inequality.

Now consider the conditioned event, $ \Pr [\exists t\in [T] \text{ s.t. } e\in \delta (S_t)|\rho_1,\ldots,\rho_T\leq 1]$.
As $ \Pr [\rho_1,\ldots,\rho_T\leq 1] = 1-\Pr [\exists t\in [T]:\rho_t >1]\geq 1-T\cdot \text{exp}(-\lambda)\geq \nicefrac{1}{2}$ (recall that $\lambda=\ln{(2T)}$), we get:
\[ \Pr [\exists t\in [T] \text{ s.t. } e\in \delta (S_t)|\rho_1,\ldots,\rho_T\leq 1] \leq \frac{\Pr [\exists t\in [T] \text{ s.t. } e\in \delta (S_t)]}{\Pr [\rho_1,\ldots,\rho_T\leq 1]}\leq \frac{2d(u,v)\cdot \ln{(2T)}}{R}. \qedhere \]
\end{proof}

\begin{proof}[Proof of Lemma \ref{lem:cutting}]
Conditioned on $\mathcal{E}$, the expected number of edges cut 
is at most $ 2\sum_{(u,v)\in E} d(u,v)\cdot \ln{(2T)}/R$ (by Lemma \ref{lem:auxCutting}).
Let $\mathcal{B}$ denote the 
event that more than $4 \sum_{(u,v)\in E} d(u,v)\cdot \ln{(2T)}/R$ are cut.
Then  $\Pr[\mathcal{\overline{B}} \cap \mathcal{E}] = \Pr[\mathcal{E}] \Pr[\mathcal{\overline{B}}| \mathcal{E}] \geq 1/4$, and thus in expectation after $4$ executions of Algorithm \ref{alg:cutting} both these properties will be satisfied.
Thus, requirement (\ref{decompose1}) holds.

Requirement (\ref{decompose2}) follows since $ \rho_t\leq 1$.

Requirement (\ref{decompose3}) follows since when $S_t$ is constructed by Algorithm \ref{alg:cutting}, all vertices in $B(v_t,R)\cap V$ (as $ \rho_t\geq 0$) that were not previously covered by $ S_1,\ldots,S_{t-1}$ must be covered.
\end{proof}

\section{Proof of Lemma  \ref{fact:NewBallBound}}\label{app:NewBallBound}
{\bf Lemma
\ref{fact:NewBallBound}.} (restated)
    For every $v$, the quantities $i_v,j_v \in [2,\ell+1]$. Moreover, $|B(v,\Delta _{i_v-1})| \leq n_{j_v-2}$, where we use the convention $n_0=n/2$ to handle $j_v=2$. In particular, $ |B(v,\Delta_{i_v-1})|\leq n/2$ for all $v$.
\begin{proof}
Notice that in Definition \ref{def:radius-scale} we require that  $i_v\geq 2$. This ensures that $\Delta_{i_v-1} \leq \Delta _1 =n/8$ for every $v$, and 
hence $|B(v,\Delta_{i_v-1})| \leq n/2$ 
by Lemma \ref{lem:size-ball}.

Now, as $|B(v,\Delta_1)| = |B(v,n/8)|\leq n/2 = n_1$ by Lemma \ref{lem:size-ball} for any $v$, and as \eqref{eq:radius-scale} does not hold for $i< i_v$, we have that
\begin{equation}
    \label{eq:ball-relation} |B(v,\Delta_i)| \leq n_i  ~~~~~~~ \forall i=1,\ldots, i_v-1.
\end{equation}
This implies that $i_v \leq \ell+1$ for all $v$ as
$ n_{\ell} \leq 1 \leq  |B_v| \leq |B(v,\Delta_{i_v-1})| \leq n_{i_v-1}$, where the last inequality uses \eqref{eq:ball-relation}.
Similarly, $j_v\geq 2$ for all $v$, as  $ n_{j_v} <|B_v|\leq n/2 = n_1$, and $j_v\leq \ell+1$ as $ n_{\ell}\leq 1 \leq |B_v| \leq n_{j_v-1}$. Thus $i_v,j_v \in [2,\ell+1]$.

We now show that  $|B(v,\Delta _{i_v-1})| \leq n_{j_v-2}$. If $j_v\geq 3$,
then as $|B_v| \leq n_{j_v-1}$ (by definition of size scale), and as $i_v$ satisfies \eqref{eq:radius-scale}, using the relation \eqref{n_sequence} between $n_{j_v-1}$ and $n_{j_v-2}$ we have that  $|B(v,\Delta_{i_v-1})| \leq n_{j_v-2}$.

For $j_v= 2$ (in fact any $j_v\geq 2)$, we have  $|B(v,\Delta _{i_v-1})| \leq |B(v,\Delta_1)| = |B(v,n/8)| \leq n/2$, where the first inequality holds as $i_v\geq 2$ and second by Lemma \ref{lem:size-ball}. 
\end{proof}

\end{document}